%% file: main.tex
\documentclass[journal]{IEEEtran}

\usepackage{bm}
\usepackage[utf8]{inputenc}
\usepackage{algorithm}
\usepackage{algpseudocode}
\usepackage{amsmath,amssymb,amsfonts, amsthm}
\usepackage{balance}
\usepackage{booktabs}
\usepackage{caption}
\usepackage{tabularx}
\usepackage{cite}
\usepackage{color}
\usepackage{comment}
\usepackage{calc}
\usepackage{float}
\usepackage{graphicx} 
\usepackage{multirow}
\usepackage{nicefrac}
\usepackage{subfigure}
\usepackage{enumerate}
\usepackage{textcomp}
\usepackage{verbatim}
\usepackage{xcolor}
\usepackage{lipsum}  
\usepackage{units}
\usepackage{subcaption}
\usepackage{orcidlink}
\usepackage{pgfplots}
\usepackage{tikz}
\usepackage{caption}
\usepackage{subcaption}
\usepackage{graphicx} 
\usepackage{booktabs} 
\usepackage{array}   
\usepackage{tabularx}
\usepackage{graphicx}

\usetikzlibrary{calc,arrows.meta,decorations.pathreplacing}

\DeclareUnicodeCharacter{2212}{−}
\pgfplotsset{width=10cm,compat=1.9}
\usepgfplotslibrary{groupplots,dateplot}
\usetikzlibrary{patterns,shapes.arrows}
\pgfplotsset{compat=newest}

\usepackage[acronyms,nonumberlist,nopostdot,nomain,nogroupskip,acronymlists={hidden}]{glossaries}

\newglossary[algh]{hidden}{acrh}{acnh}{Hidden Acronyms}

\input{acronyms.tex}

\begin{document}

\title{Analyzing Zigbee Traffic: Datasets, Classification and Storage Trade-offs}
\author{Antonio Boiano,
        Dalin Zheng,
        Fabio Palmese,
        Andrea Pimpinella
        and~Alessandro~E.~C.~Redondi\
\thanks{A. Boiano (antonio.boiano@polimi.it), F. Palmese (fabio.palmese@polimi.it) and A. E. C. Redondi ( alessandroenrico.redondi@polimi.it), are with the Department of Electronics, Information and Bioengineering, Politecnico di Milano, Milan, Italy. Andrea Pimpinella (andrea.pimpinella@unibg.it) is with the Department of Management, Information and Production Engineering (DIGIP), University of Bergamo, Bergamo, Italy. Dalin Zhang (924453869@stu.xjtu.edu.cn) is with the School of Computer Science and Technology, Shaanxi Province Key Laboratory of Computer Network, Xi’an Jiaotong University, Xi’an, 710049, China}%
}


\markboth{Journal of \LaTeX\ Class Files,~Vol.~14, No.~8, August~2015}%
{Shell \MakeLowercase{\textit{et al.}}: Bare Demo of IEEEtran.cls for IEEE Journals}

\maketitle

\begin{abstract}
Zigbee is widely used in smart home environments due to its low power consumption and support for mesh networking, making it a relevant target for traffic-based IoT forensic analysis. However, existing studies often rely on limited datasets and fixed network configurations.
In this paper, we analyze Zigbee network traffic from three complementary perspectives: data collection, traffic classification, and storage efficiency. We introduce ZIOTP2025, a publicly available dataset of Zigbee traffic collected from commercial smart home devices deployed under multiple network configurations and capturing realistic interaction scenarios.
Using this dataset, we study two traffic classification tasks: device type classification and individual device identification, and evaluate their robustness under both intra-configuration and cross-configuration settings. Our results show that while high classification accuracy can be achieved under controlled conditions, performance degrades significantly when models are evaluated across different network configurations, particularly for fine-grained identification tasks.
Finally, we investigate the trade-off between traffic storage requirements and classification accuracy. We show that lossy compression of traffic features through quantization can reduce storage requirements by approximately 4--5$\times$ compared to lossless storage of raw packet traces, while preserving near-lossless classification performance.
Overall, our results highlight the need for topology-aware Zigbee traffic analysis and storage-efficient feature compression to enable robust and scalable IoT forensic systems.

\end{abstract}


\IEEEpeerreviewmaketitle

%
%
%
%

\input{Chapters/introduction}
\input{Chapters/background_v2}
\input{Chapters/sota_v2}
\input{Chapters/dataset_v2}
\input{Chapters/methodology_v3}
\input{Chapters/results_v3}

\input{Chapters/compression_v3}
\input{Chapters/conclusion_v2}


%

\section*{Acknowledgment}
This study was carried out within the PRIN project COMPACT and received funding from Next Generation EU, Mission 4 Component 1, CUP: D53D23001340006.

\ifCLASSOPTIONcaptionsoff
  \newpage
\fi



%

\bibliographystyle{IEEEtran}
\bibliography{main}

%

\begin{IEEEbiography}[{\includegraphics[width=1in,height=1.25in,clip,keepaspectratio]{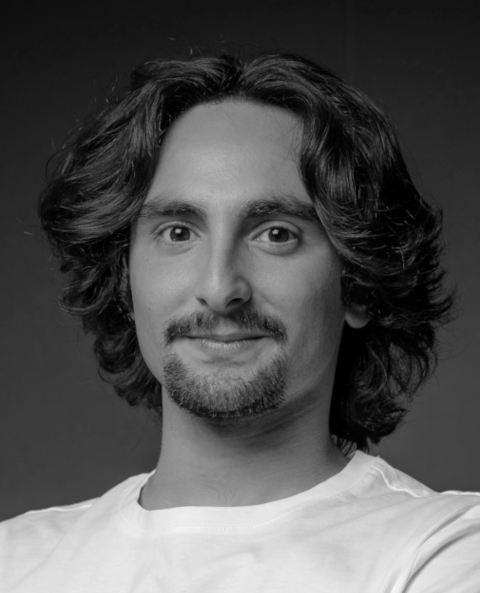}}]{Antonio Boiano (Graduate Student Member, IEEE)}
received the M.Sc. degree in automation and control engineering from the Politecnico di Milano, Milan, Italy, in 2021, where he is currently pursuing the Ph.D. degree in telecommunications with the Advanced Network Technologies Laboratory, at the Department of Electronics, Information and Bioengineering (DEIB), Politecnico di Milano. His research activities focus on network traffic analysis with major focus on Federated Learning, IoT and mobile traffic analysis.
\end{IEEEbiography}
\begin{IEEEbiography}[{\includegraphics[width=1in,height=1.25in,clip,keepaspectratio]{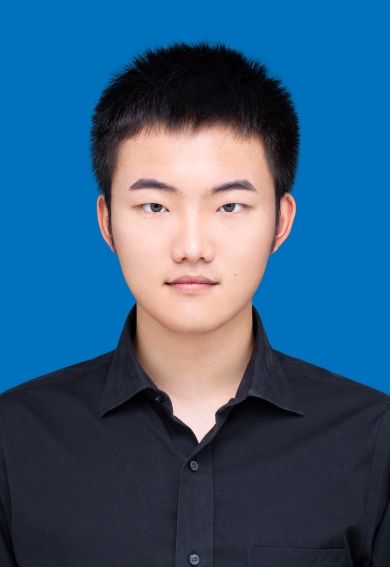}}]{Dalin Zheng}
received his bachelor’s degree from Xi’an Jiaotong University, Xi’an, China, in 2023. He pursued his master’s studies at the same university from 2023 to 2025, and is currently pursuing his doctoral degree at the Shaanxi Province Key Laboratory of Computer Network, Xi’an Jiaotong University. His research interests include data aggregation, data security, IoT information security, federate learning and IoT forensics.
\end{IEEEbiography}
\begin{IEEEbiography}[{\includegraphics[width=1in,height=1.25in,clip,keepaspectratio]{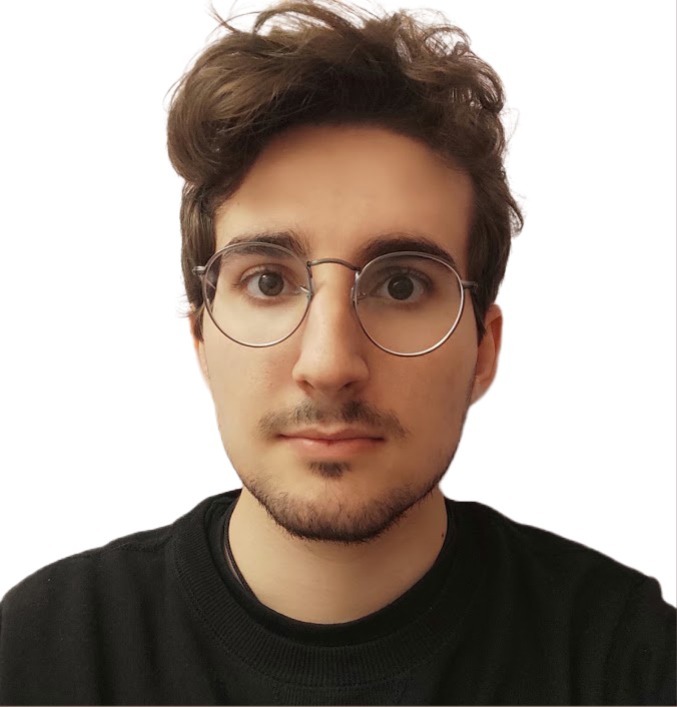}}]{Fabio Palmese (Member, IEEE))}
 received the M.Sc. degree in computer science and engineering and the Ph.D. degree in information technology in the telecommunications area from the Politecnico di Milano, Milan, Italy, in 2020 and 2025, respectively. From September 2023 to March 2024, he was a visiting student with the EEE Department, University College London, U.K. His research activities focus on network traffic analysis and Internet of Things topics. His Ph.D. research thesis focused on the IoT forensics, with specific attention to IoT network traffic collection and analysis for forensic investigations.
\end{IEEEbiography}
\begin{IEEEbiography}
[{\includegraphics[width=1in,height=1.25in,clip,keepaspectratio]{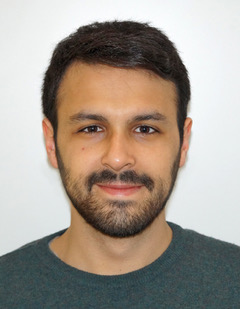}}]{Andrea Pimpinella} is currently a Researcher at the Management, Information and Production Engineering of Università degli Studi di Bergamo, Italy. Andrea received in February 2022 a PhD Title Cum Laude in Information Technology from Politecnico di Milano. Andrea was also a visiting scholar at the department of Urban Studies and Planning of MIT, Boston, from January to March 2023, and is currently in the editorial board of Elsevier's Computer Communications. His research activities are focused on the development of data driven, QoE oriented and artificial intelligence based approaches to monitoring and management of communication networks, including: cellular traffic forecasting, QoE oriented anomaly detection in mobile networks, energy-aware networking and performance optimization in IoT networks.
\end{IEEEbiography}
\begin{IEEEbiography}
[{\includegraphics[width=1in,height=1.25in,clip,keepaspectratio]{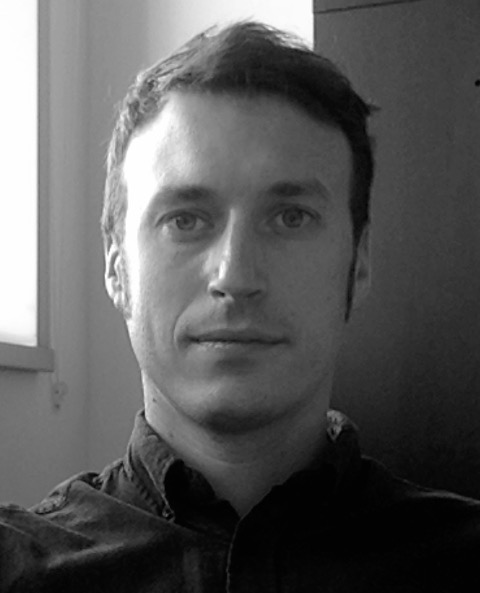}}]{Alessandro E. C. Redondi (Senior Member, IEEE)} is Associate Professor with the Dipartimento di Elettronica, Informazione e Bioingegneria of the Politecnico di Milano, Italy. He received the MS in Computer Engineering in July 2009 and the Ph.D. in Information Engineering in February 2014, both from Politecnico di Milano. From September 2012 to April 2013 he was a visiting student at the EEE Department of the University College of London (UCL). In 2018 he was a vitising researcher at the Computer Network Research Group of Universitat Politecnica de Valencia. His research activities are focused on the design and optimization of wireless/IoT systems and on network data analytics.
\end{IEEEbiography}




\end{document}

%% file: acronyms.tex
\newacronym{3gpp}{3GPP}{3rd Generation Partnership Project}
\newacronym{REST}{REST}{Representational State Transfer}
\newacronym{MQTT}{MQTT}{Message Queuing Telemetry Transport}
\newacronym{AMQP}{AMQP}{Advanced Message Queuing Protocol}
\newacronym{TCP}{TCP}{Transmission Control Protocol}
\newacronym{CoAP}{CoAP}{Constrained Application Protocol}
\newacronym{FedAvg}{FedAvg}{Federated Averaging}
\newacronym{MLP}{MLP}{Multilayer Perceptron }
\newacronym{gRPC}{gRPC}{Google Remote Procedure Calls}
\newacronym{MNIST}{MNIST}{Modified National Institute of Standards and Technology database}
\newacronym{IID}{IID}{Independent and Identically Distributed}
\newacronym{Non-IID}{Non-IID}{Non-Independent and Non-Identically Distributed}
\newacronym{FL}{FL}{Federated Learning}
\newacronym{4g}{4G}{4th generation}
\newacronym{Wi-Fi}{Wi-Fi}{Wireless Fidelity}
\newacronym{soa}{SoA}{State of the Art}
\newacronym{smtp}{SMTP}{Simple Mail Transfer Protocol}

\newacronym{ciot}{CIoT}{Consumer IoT}

\newacronym{5g}{5G}{5th generation}
\newacronym{6g}{6G}{6th generation}
\newacronym{5gc}{5GC}{5G Core}
\newacronym{adc}{ADC}{Analog to Digital Converter}
\newacronym{aerpaw}{AERPAW}{Aerial Experimentation and Research Platform for Advanced Wireless}
\newacronym{ai}{AI}{Artificial Intelligence}
\newacronym{aimd}{AIMD}{Additive Increase Multiplicative Decrease}
\newacronym{am}{AM}{Acknowledged Mode}
\newacronym{amc}{AMC}{Adaptive Modulation and Coding}
\newacronym{amf}{AMF}{Access and Mobility Management Function}
\newacronym{aops}{AOPS}{Adaptive Order Prediction Scheduling}
\newacronym{api}{API}{Application Programming Interface}
\newacronym{apn}{APN}{Access Point Name}
\newacronym{ap}{AP}{Application Protocol}
\newacronym{aqm}{AQM}{Active Queue Management}
\newacronym{ar}{AR}{Augmented Reality}
\newacronym{ausf}{AUSF}{Authentication Server Function}
\newacronym{avc}{AVC}{Advanced Video Coding}
\newacronym{awgn}{AGWN}{Additive White Gaussian Noise}
\newacronym{balia}{BALIA}{Balanced Link Adaptation Algorithm}
\newacronym{bbu}{BBU}{Base Band Unit}
\newacronym{bdp}{BDP}{Bandwidth-Delay Product}
\newacronym{ber}{BER}{Bit Error Rate}
\newacronym{bf}{BF}{Beamforming}
\newacronym{bler}{BLER}{Block Error Rate}
\newacronym{brr}{BRR}{Bayesian Ridge Regressor}
\newacronym{bs}{BS}{Base Station}
\newacronym{bsr}{BSR}{Buffer Status Report}
\newacronym{bss}{BSS}{Business Support System}
\newacronym{ca}{CA}{Carrier Aggregation}
\newacronym{caas}{CaaS}{Connectivity-as-a-Service}
\newacronym{cb}{CB}{Code Block}
\newacronym{cc}{CC}{Congestion Control}
\newacronym{ccid}{CCID}{Congestion Control ID}
\newacronym{cco}{CC}{Carrier Component}
\newacronym{cdd}{CDD}{Cyclic Delay Diversity}
\newacronym{cdf}{CDF}{Cumulative Distribution Function}
\newacronym{cdn}{CDN}{Content Distribution Network}
\newacronym{cn}{CN}{Core Network}
\newacronym{codel}{CoDel}{Controlled Delay Management}
\newacronym{comac}{COMAC}{Converged Multi-Access and Core}
\newacronym{cord}{CORD}{Central Office Re-architected as a Datacenter}
\newacronym{cornet}{CORNET}{COgnitive Radio NETwork}
\newacronym{cosmos}{COSMOS}{Cloud Enhanced Open Software Defined Mobile Wireless Testbed for City-Scale Deployment}
\newacronym{cots}{COTS}{Commercial Off-the-Shelf}
\newacronym{cp}{CP}{Control Plane}
\newacronym{cyp}{CP}{Cyclic Prefix}
\newacronym{up}{UP}{User Plane}
\newacronym{cpu}{CPU}{Central Processing Unit}
\newacronym{cqi}{CQI}{Channel Quality Information}
\newacronym{cr}{CR}{Cognitive Radio}
\newacronym{cran}{C-RAN}{Cloud \gls{ran}}
\newacronym{crs}{CRS}{Cell Reference Signal}
\newacronym{csi}{CSI}{Channel State Information}
\newacronym{csirs}{CSI-RS}{Channel State Information - Reference Signal}
\newacronym{cu}{CU}{Central Unit}
\newacronym{d2tcp}{D$^2$TCP}{Deadline-aware Data center TCP}
\newacronym{d3}{D$^3$}{Deadline-Driven Delivery}
\newacronym{dac}{DAC}{Digital to Analog Converter}
\newacronym{dag}{DAG}{Directed Acyclic Graph}
\newacronym{das}{DAS}{Distributed Antenna System}
\newacronym{dash}{DASH}{Dynamic Adaptive Streaming over HTTP}
\newacronym{dc}{DC}{Dual Connectivity}
\newacronym{dccp}{DCCP}{Datagram Congestion Control Protocol}
\newacronym{dce}{DCE}{Direct Code Execution}
\newacronym{dci}{DCI}{Downlink Control Information}
\newacronym{dctcp}{DCTCP}{Data Center TCP}
\newacronym{dl}{DL}{Downlink}
\newacronym{dmr}{DMR}{Deadline Miss Ratio}
\newacronym{dmrs}{DMRS}{DeModulation Reference Signal}
\newacronym{drlcc}{DRL-CC}{Deep Reinforcement Learning Congestion Control}
\newacronym{drs}{DRS}{Discovery Reference Signal}
\newacronym{du}{DU}{Distributed Unit}
\newacronym{e2e}{E2E}{end-to-end}
\newacronym{ecaas}{ECaaS}{Edge-Cloud-as-a-Service}
\newacronym{ecn}{ECN}{Explicit Congestion Notification}
\newacronym{edf}{EDF}{Earliest Deadline First}
\newacronym{embb}{eMBB}{Enhanced Mobile Broadband}
\newacronym{empower}{EMPOWER}{EMpowering transatlantic PlatfOrms for advanced WirEless Research}
\newacronym{enb}{eNB}{evolved Node Base}
\newacronym{endc}{EN-DC}{E-UTRAN-\gls{nr} \gls{dc}}
\newacronym{epc}{EPC}{Evolved Packet Core}
\newacronym{eps}{EPS}{Evolved Packet System}
\newacronym{es}{ES}{Edge Server}
\newacronym{etsi}{ETSI}{European Telecommunications Standards Institute}
\newacronym[firstplural=Estimated Times of Arrival (ETAs)]{eta}{ETA}{Estimated Time of Arrival}
\newacronym{eutran}{E-UTRAN}{Evolved Universal Terrestrial Access Network}
\newacronym{faas}{FaaS}{Function-as-a-Service}
\newacronym{fapi}{FAPI}{Functional Application Platform Interface}
\newacronym{fdd}{FDD}{Frequency Division Duplexing}
\newacronym{fdm}{FDM}{Frequency Division Multiplexing}
\newacronym{fdma}{FDMA}{Frequency Division Multiple Access}
\newacronym{fed4fire}{FED4FIRE+}{Federation 4 Future Internet Research and Experimentation Plus}
\newacronym{fir}{FIR}{Finite Impulse Response}
\newacronym{fit}{FIT}{Future \acrlong{iot}}
\newacronym{fpga}{FPGA}{Field Programmable Gate Array}
\newacronym{fr2}{FR2}{Frequency Range 2}
\newacronym{fs}{FS}{Fast Switching}
\newacronym{fscc}{FSCC}{Flow Sharing Congestion Control}
\newacronym{ftp}{FTP}{File Transfer Protocol}
\newacronym{fw}{FW}{Flow Window}
\newacronym{ge}{GE}{Gaussian Elimination}
\newacronym{gnb}{gNB}{Next Generation Node Base}
\newacronym{gop}{GOP}{Group of Pictures}
\newacronym{gpr}{GPR}{Gaussian Process Regressor}
\newacronym{gpu}{GPU}{Graphics Processing Unit}
\newacronym{gtp}{GTP}{GPRS Tunneling Protocol}
\newacronym{gtpc}{GTP-C}{GPRS Tunnelling Protocol Control Plane}
\newacronym{gtpu}{GTP-U}{GPRS Tunnelling Protocol User Plane}
\newacronym{gtpv2c}{GTPv2-C}{\gls{gtp} v2 - Control}
\newacronym{gw}{GW}{Gateway}
\newacronym{harq}{HARQ}{Hybrid Automatic Repeat reQuest}
\newacronym{hetnet}{HetNet}{Heterogeneous Network}
\newacronym{hh}{HH}{Hard Handover}
\newacronym{hol}{HOL}{Head-of-Line}
\newacronym{hqf}{HQF}{Highest-quality-first}
\newacronym{hss}{HSS}{Home Subscription Server}
\newacronym{http}{HTTP}{HyperText Transfer Protocol}
\newacronym{ia}{IA}{Initial Access}
\newacronym{iab}{IAB}{Integrated Access and Backhaul}
\newacronym{ic}{IC}{Incident Command}
\newacronym{ietf}{IETF}{Internet Engineering Task Force}
\newacronym{imsi}{IMSI}{International Mobile Subscriber Identity}
\newacronym{imt}{IMT}{International Mobile Telecommunication}
\newacronym{iot}{IoT}{Internet of Things}
\newacronym{ip}{IP}{Internet Protocol}
\newacronym{itu}{ITU}{International Telecommunication Union}
\newacronym{kpi}{KPI}{Key Performance Indicator}
\newacronym{kpm}{KPM}{Key Performance Measurement}
\newacronym{kvm}{KVM}{Kernel-based Virtual Machine}
\newacronym{los}{LOS}{Line-of-Sight}
\newacronym{lsm}{LSM}{Link-to-System Mapping}
\newacronym{lstm}{LSTM}{Long Short Term Memory}
\newacronym{lte}{LTE}{Long Term Evolution}
\newacronym{lxc}{LXC}{Linux Container}
\newacronym{m2m}{M2M}{Machine to Machine}
\newacronym{mac}{MAC}{Medium Access Control}
\newacronym{manet}{MANET}{Mobile Ad Hoc Network}
\newacronym{mano}{MANO}{Management and Orchestration}
\newacronym{mc}{MC}{Multi-Connectivity}
\newacronym{mcc}{MCC}{Mobile Cloud Computing}
\newacronym{mchem}{MCHEM}{Massive Channel Emulator}
\newacronym{mcs}{MCS}{Modulation and Coding Scheme}
\newacronym{mec}{MEC}{Multi-access Edge Computing}
\newacronym{mec2}{MEC}{Mobile Edge Cloud}
\newacronym{mfc}{MFC}{Mobile Fog Computing}
\newacronym{mgen}{MGEN}{Multi-Generator}
\newacronym{mi}{MI}{Mutual Information}
\newacronym{mib}{MIB}{Master Information Block}
\newacronym{miesm}{MIESM}{Mutual Information Based Effective SINR}
\newacronym{mimo}{MIMO}{Multiple Input, Multiple Output}
\newacronym{ml}{ML}{Machine Learning}
\newacronym{mlr}{MLR}{Maximum-local-rate}
\newacronym[plural=\gls{mme}s,firstplural=Mobility Management Entities (MMEs)]{mme}{MME}{Mobility Management Entity}
\newacronym{mmtc}{mMTC}{Massive Machine-Type Communications}
\newacronym{mmwave}{mmWave}{millimeter wave}
\newacronym{mpdccp}{MP-DCCP}{Multipath Datagram Congestion Control Protocol}
\newacronym{mptcp}{MPTCP}{Multipath TCP}
\newacronym{mr}{MR}{Maximum Rate}
\newacronym{mrdc}{MR-DC}{Multi \gls{rat} \gls{dc}}
\newacronym{mse}{MSE}{Mean Square Error}
\newacronym{mss}{MSS}{Maximum Segment Size}
\newacronym{mt}{MT}{Mobile Termination}
\newacronym{mtd}{MTD}{Machine-Type Device}
\newacronym{mtu}{MTU}{Maximum Transmission Unit}
\newacronym{mumimo}{MU-MIMO}{Multi-user \gls{mimo}}
\newacronym{mvno}{MVNO}{Mobile Virtual Network Operator}
\newacronym{nalu}{NALU}{Network Abstraction Layer Unit}
\newacronym{nas}{NAS}{Non-Access Stratum}
\newacronym{nbiot}{NB-IoT}{Narrow Band IoT}
\newacronym{nfv}{NFV}{Network Function Virtualization}
\newacronym{nfvi}{NFVI}{Network Function Virtualization Infrastructure}
\newacronym{ngrg}{nGRG}{next Generation Research Group}
\newacronym{ni}{NI}{Network Interfaces}
\newacronym{nic}{NIC}{Network Interface Card}
\newacronym{nlos}{NLOS}{Non-Line-of-Sight}
\newacronym{now}{NOW}{Non Overlapping Window}
\newacronym{nsm}{NSM}{Network Service Mesh}
\newacronym{nr}{NR}{New Radio}
\newacronym{nrf}{NRF}{Network Repository Function}
\newacronym{nsa}{NSA}{Non Stand Alone}
\newacronym{nse}{NSE}{Network Slicing Engine}
\newacronym{nssf}{NSSF}{Network Slice Selection Function}
\newacronym{o2i}{O2I}{Outdoor to Indoor}
\newacronym{oai}{OAI}{OpenAirInterface}
\newacronym{oaicn}{OAI-CN}{\gls{oai} \acrlong{cn}}
\newacronym{oairan}{OAI-RAN}{\acrlong{oai} \acrlong{ran}}
\newacronym{oam}{OAM}{Operations, Administration and Maintenance}
\newacronym{ofdm}{OFDM}{Orthogonal Frequency Division Multiplexing}
\newacronym{olia}{OLIA}{Opportunistic Linked Increase Algorithm}
\newacronym{omec}{OMEC}{Open Mobile Evolved Core}
\newacronym{onap}{ONAP}{Open Network Automation Platform}
\newacronym{onf}{ONF}{Open Networking Foundation}
\newacronym{onos}{ONOS}{Open Networking Operating System}
\newacronym{oom}{OOM}{\gls{onap} Operations Manager}
\newacronym{opnfv}{OPNFV}{Open Platform for \gls{nfv}}
\newacronym{oran}{O-RAN}{Open Radio Access Network}
\newacronym{orbit}{ORBIT}{Open-Access Research Testbed for Next-Generation Wireless Networks}
\newacronym{os}{OS}{Operating System}
\newacronym{oss}{OSS}{Operations Support System}
\newacronym{otic}{OTIC}{Open Testing \& Integration Centre}
\newacronym{pa}{PA}{Position-aware}
\newacronym{pase}{PASE}{Prioritization, Arbitration, and Self-adjusting Endpoints}
\newacronym{pawr}{PAWR}{Platforms for Advanced Wireless Research}
\newacronym{pbch}{PBCH}{Physical Broadcast Channel}
\newacronym{pcef}{PCEF}{Policy and Charging Enforcement Function}
\newacronym{pcfich}{PCFICH}{Physical Control Format Indicator Channel}
\newacronym{pcrf}{PCRF}{Policy and Charging Rules Function}
\newacronym{pdcch}{PDCCH}{Physical Downlink Control Channel}
\newacronym{pdcp}{PDCP}{Packet Data Convergence Protocol}
\newacronym{pdsch}{PDSCH}{Physical Downlink Shared Channel}
\newacronym{pdu}{PDU}{Packet Data Unit}
\newacronym{pf}{PF}{Proportional Fair}
\newacronym{pgw}{PGW}{Packet Gateway}
\newacronym{phich}{PHICH}{Physical Hybrid ARQ Indicator Channel}
\newacronym{phy}{PHY}{Physical}
\newacronym{pmch}{PMCH}{Physical Multicast Channel}
\newacronym{pmi}{PMI}{Precoding Matrix Indicators}
\newacronym{powder}{POWDER}{Platform for Open Wireless Data-driven Experimental Research}
\newacronym{ppo}{PPO}{Proximal Policy Optimization}
\newacronym{ppp}{PPP}{Poisson Point Process}
\newacronym{prach}{PRACH}{Physical Random Access Channel}
\newacronym{prb}{PRB}{Physical Resource Block}
\newacronym{psnr}{PSNR}{Peak Signal to Noise Ratio}
\newacronym{pss}{PSS}{Primary Synchronization Signal}
\newacronym{pucch}{PUCCH}{Physical Uplink Control Channel}
\newacronym{pusch}{PUSCH}{Physical Uplink Shared Channel}
\newacronym{qam}{QAM}{Quadrature Amplitude Modulation}
\newacronym{qci}{QCI}{\gls{qos} Class Identifier}
\newacronym{qoe}{QoE}{Quality of Experience}
\newacronym{qos}{QoS}{Quality of Service}
\newacronym{quic}{QUIC}{Quick UDP Internet Connections}
\newacronym{rach}{RACH}{Random Access Channel}
\newacronym{ran}{RAN}{Radio Access Network}
\newacronym[firstplural=Radio Access Technologies (RATs)]{rat}{RAT}{Radio Access Technology}
\newacronym{rcn}{RCN}{Research Coordination Network}
\newacronym{rc}{RC}{RAN Control}
\newacronym{rec}{REC}{Radio Edge Cloud}
\newacronym{red}{RED}{Random Early Detection}
\newacronym{renew}{RENEW}{Reconfigurable Eco-system for Next-generation End-to-end Wireless}
\newacronym{rf}{RF}{Random Forest}
\newacronym{rfc}{RFC}{Request for Comments}
\newacronym{rfr}{RFR}{Random Forest Regressor}
\newacronym{ric}{RIC}{\gls{ran} Intelligent Controller}
\newacronym{rlc}{RLC}{Radio Link Control}
\newacronym{rlf}{RLF}{Radio Link Failure}
\newacronym{rlnc}{RLNC}{Random Linear Network Coding}
\newacronym{rmr}{RMR}{RIC Message Router}
\newacronym{rmse}{RMSE}{Root Mean Squared Error}
\newacronym{rnis}{RNIS}{Radio Network Information Service}
\newacronym{rr}{RR}{Round Robin}
\newacronym{rrc}{RRC}{Radio Resource Control}
\newacronym{rrm}{RRM}{Radio Resource Management}
\newacronym{rru}{RRU}{Remote Radio Unit}
\newacronym{rs}{RS}{Remote Server}
\newacronym{rsrp}{RSRP}{Reference Signal Received Power}
\newacronym{rsrq}{RSRQ}{Reference Signal Received Quality}
\newacronym{rss}{RSS}{Received Signal Strength}
\newacronym{rssi}{RSSI}{Received Signal Strength Indicator}
\newacronym{rtt}{RTT}{Round Trip Time}
\newacronym{ru}{RU}{Radio Unit}
\newacronym{rw}{RW}{Receive Window}
\newacronym{rx}{RX}{Receiver}
\newacronym{s1ap}{S1AP}{S1 Application Protocol}
\newacronym{sa}{SA}{standalone}
\newacronym{sack}{SACK}{Selective Acknowledgment}
\newacronym{sap}{SAP}{Service Access Point}
\newacronym{sc2}{SC2}{Spectrum Collaboration Challenge}
\newacronym{scef}{SCEF}{Service Capability Exposure Function}
\newacronym{sch}{SCH}{Secondary Cell Handover}
\newacronym{scoot}{SCOOT}{Split Cycle Offset Optimization Technique}
\newacronym{sctp}{SCTP}{Stream Control Transmission Protocol}
\newacronym{sdap}{SDAP}{Service Data Adaptation Protocol}
\newacronym{sdk}{SDK}{Software Development Kit}
\newacronym{sdm}{SDM}{Space Division Multiplexing}
\newacronym{sdma}{SDMA}{Spatial Division Multiple Access}
\newacronym{sdn}{SDN}{Software-defined Networking}
\newacronym{sdr}{SDR}{Software-defined Radio}
\newacronym{seba}{SEBA}{SDN-Enabled Broadband Access}
\newacronym{sgsn}{SGSN}{Serving GPRS Support Node}
\newacronym{sgw}{SGW}{Service Gateway}
\newacronym{si}{SI}{Study Item}
\newacronym{sib}{SIB}{Secondary Information Block}
\newacronym{sinr}{SINR}{Signal to Interference plus Noise Ratio}
\newacronym{sip}{SIP}{Session Initiation Protocol}
\newacronym{siso}{SISO}{Single Input, Single Output}
\newacronym{sla}{SLA}{Service Level Agreement}
\newacronym{sm}{SM}{Service Model}
\newacronym{smf}{SMF}{Session Management Function}
\newacronym{smo}{SMO}{Service Management and Orchestration}
\newacronym{sms}{SMS}{Short Message Service}
\newacronym{smsgmsc}{SMS-GMSC}{\gls{sms}-Gateway}
\newacronym{snr}{SNR}{Signal-to-Noise-Ratio}
\newacronym{son}{SON}{Self-Organizing Network}
\newacronym{sptcp}{SPTCP}{Single Path TCP}
\newacronym{srb}{SRB}{Service Radio Bearer}
\newacronym{srn}{SRN}{Standard Radio Node}
\newacronym{srs}{SRS}{Sounding Reference Signal}
\newacronym{ss}{SS}{Synchronization Signal}
\newacronym{sss}{SSS}{Secondary Synchronization Signal}
\newacronym{st}{ST}{Spanning Tree}
\newacronym{svc}{SVC}{Scalable Video Coding}
\newacronym{tb}{TB}{Transport Block}
\newacronym{tcp}{TCP}{Transmission Control Protocol}
\newacronym{tdd}{TDD}{Time Division Duplexing}
\newacronym{tdm}{TDM}{Time Division Multiplexing}
\newacronym{tdma}{TDMA}{Time Division Multiple Access}
\newacronym{tfl}{TfL}{Transport for London}
\newacronym{tfrc}{TFRC}{TCP-Friendly Rate Control}
\newacronym{tft}{TFT}{Traffic Flow Template}
\newacronym{tgen}{TGEN}{Traffic Generator}
\newacronym{tip}{TIP}{Telecom Infra Project}
\newacronym{tm}{TM}{Transparent Mode}
\newacronym{to}{TO}{Telco Operator}
\newacronym{tr}{TR}{Technical Report}
\newacronym{trp}{TRP}{Transmitter Receiver Pair}
\newacronym{ts}{TS}{Technical Specification}
\newacronym{tti}{TTI}{Transmission Time Interval}
\newacronym{ttt}{TTT}{Time-to-Trigger}
\newacronym{tx}{TX}{Transmitter}
\newacronym{uas}{UAS}{Unmanned Aerial System}
\newacronym{uav}{UAV}{Unmanned Aerial Vehicle}
\newacronym{udm}{UDM}{Unified Data Management}
\newacronym{udp}{UDP}{User Datagram Protocol}
\newacronym{udr}{UDR}{Unified Data Repository}
\newacronym{ue}{UE}{User Equipment}
\newacronym{uhd}{UHD}{\gls{usrp} Hardware Driver}
\newacronym{ul}{UL}{Uplink}
\newacronym{um}{UM}{Unacknowledged Mode}
\newacronym{uml}{UML}{Unified Modeling Language}
\newacronym{upa}{UPA}{Uniform Planar Array}
\newacronym{upf}{UPF}{User Plane Function}
\newacronym{urllc}{URLLC}{Ultra Reliable and Low Latency Communications}
\newacronym{usa}{U.S.}{United States}
\newacronym{usim}{USIM}{Universal Subscriber Identity Module}
\newacronym{usrp}{USRP}{Universal Software Radio Peripheral}
\newacronym{utc}{UTC}{Urban Traffic Control}
\newacronym{vim}{VIM}{Virtualization Infrastructure Manager}
\newacronym{vm}{VM}{Virtual Machine}
\newacronym{vnf}{VNF}{Virtual Network Function}
\newacronym{volte}{VoLTE}{Voice over \gls{lte}}
\newacronym{voltha}{VOLTHA}{Virtual OLT HArdware Abstraction}
\newacronym{vr}{VR}{Virtual Reality}
\newacronym{vran}{vRAN}{Virtualized \gls{ran}}
\newacronym{vss}{VSS}{Video Streaming Server}
\newacronym{wbf}{WBF}{Wired Bias Function}
\newacronym{wf}{WF}{Waterfilling}
\newacronym{wg}{WG}{Working Group}
\newacronym{wlan}{WLAN}{Wireless Local Area Network}
\newacronym{osm}{OSM}{Open Source Management and Orchestration}
\newacronym{pnf}{PNF}{Physical Network Function}
\newacronym{drl}{DRL}{Deep Reinforcement Learning}
\newacronym{mtc}{MTC}{Machine-type Communications}
\newacronym{osc}{OSC}{O-RAN Software Community}
\newacronym{mns}{MnS}{Management Services}
\newacronym{ves}{VES}{\gls{vnf} Event Stream}
\newacronym{ei}{EI}{Enrichment Information}
\newacronym{fh}{FH}{Fronthaul}
\newacronym{fft}{FFT}{Fast Fourier Transform}
\newacronym{laa}{LAA}{Licensed-Assisted Access}
\newacronym{plfs}{PLFS}{Physical Layer Frequency Signals}
\newacronym{ptp}{PTP}{Precision Time Protocol}
\newacronym{asic}{ASIC}{Application-specific Integrated Circuit}
\newacronym{aal}{AAL}{Acceleration Abstraction Layer}
\newacronym{fec}{FEC}{Forward Error Correction}
\newacronym{sdl}{SDL}{Shared Data Layer}
\newacronym{nib}{NIB}{Network Information Base}
\newacronym{rnib}{R-NIB}{RAN \gls{nib}}
\newacronym{fcaps}{FCAPS}{Fault, Configuration, Accounting, Performance, Security}
\newacronym{ie}{IE}{Information Element}
\newacronym{fg}{FG}{Focus Group}
\newacronym{osfg}{OSFG}{Open Source Focus Group}
\newacronym{sdfg}{SDFG}{Standard Development Focus Group}
\newacronym{tifg}{TIFG}{Test \& Integration Focus Group}
\newacronym{sfg}{SFG}{Security Focus Group}
\newacronym{swg}{SWG}{Security Work Group}
\newacronym{e2sm}{E2SM}{E2 Service Model}
\newacronym{tsc}{TSC}{Technical Steering Committee}
\newacronym{sdo}{SDO}{Standard-Development Organization}
\newacronym{sql}{SQL}{Structured Query Language}
\newacronym{ssh}{SSH}{Secure Shell}
\newacronym{tls}{TLS}{Transport Layer Security}
\newacronym{netconf}{NETCONF}{Network Configuration Protocol}
\newacronym{dtls}{DTLS}{Datagram Transport Layer Security}
\newacronym{cmp}{CMP}{Certificate Management Protocol}
\newacronym{ccc}{CCC}{Cell Configuration and Control}
\newacronym{dsp}{DSP}{Digital Signal Processing}
\newacronym{opex}{OPEX}{Operational Expenses}
\newacronym{cbrs}{CBRS}{Citizen Broadband Radio Service}
\newacronym{ntn}{NTN}{Non-terrestrial Network}
\newacronym{gbr}{GBR}{Guaranteed Bitrate}
\newacronym{sps}{SPS}{Semi-Persistent Scheduling}
\newacronym{tbs}{TBS}{Transport Block Size}
\newacronym{gnss}{GNSS}{Global Navigation Satellite System}
\newacronym{tof}{ToF}{Time of Flight}
\newacronym{rsig}{RS}{Reference Signal}
\newacronym{nrtric}{near-RT RIC}{near-Real Time Ran Intelligent Controller}
\newacronym{aoa}{AoA}{Angle of Arrival}
\newacronym{ecdf}{ECDF}{Empirical Cumulative Distribution Function}
\newacronym{npusch}{NPUSCH}{Narrowband Physical Uplink Shared Channel}
\newacronym{mpdcch}{MPDCCH}{MTC Physical Downlink Control Channel}
\newacronym{ble}{BLE}{Bluetooth Low Energy}
\newacronym{wpan}{WPAN}{Wireless Personal Area Network}
\newacronym{pan}{PAN}{Personal Area Network}
\newacronym{lrwpan}{LR-WPAN}{Low-Rate Wireless Personal Area Network}
\newacronym{panc}{PANC}{PAN Coordinator}
\newacronym{zha}{ZHA}{Zigbee Home Automation}
\newacronym{lan}{LAN}{Local Area Network}
\newacronym{fcs}{FCS}{Frame Check Sequence}
\newacronym{fcf}{FCF}{Frame Control Field}
\newacronym{iat}{IAT}{Inter Arrival Time}
\newacronym{ack}{ACK}{Acknowledgement}
\newacronym{bilstm}{BiLSTM}{Bidirectional Long Short-Term Memory}
\newacronym{rnn}{RNN}{Recurrent Neural Network}
\newacronym{nn}{NN}{Neural Network}
\newacronym{relu}{ReLU}{Rectified Linear Unit}
\newacronym{redfd}{RFD}{Reduced Function Device}
\newacronym{ffd}{FFD}{Full Function Device}

%% file: Chapters/introduction.tex
\section{Introduction}

The Internet of Things (IoT) has become an integral component of contemporary society, with Consumer IoT (CIoT) devices accounting for nearly 50\% of the overall IoT market \cite{cisco2020cisco}. Although these technologies provide substantial benefits in everyday contexts, their pervasive deployment across domestic, professional, and public environments—together with their persistent connection to network infrastructures—raises significant concerns related to privacy and security. In this context, smart home devices are of particular interest, as they continuously and often unobtrusively collect data related to users’ daily activities. Such characteristics make them a valuable source for IoT forensics, enabling the acquisition and analysis of device-generated data, including network traffic, which can be leveraged to infer events, behaviors, or usage patterns \cite{stoyanova}. Two communication protocols are predominantly used for interconnecting smart home devices: Wi-Fi and Zigbee. While a substantial body of literature has demonstrated the feasibility of IoT forensic tools for capturing and analyzing traffic generated by Wi-Fi–enabled devices \cite{fabio,fabio2,fabio3}, forensic methodologies specifically targeting Zigbee-based devices remain comparatively underexplored. This gap is largely due to the inherent challenges associated with Zigbee traffic analysis. First, the characteristics of the Zigbee protocol, which departs significantly from the conventional TCP/IP stack, introduce additional complexity in the interpretation of captured traffic. Furthermore, Zigbee networks typically generate smaller packet sizes and lower traffic volumes than Wi-Fi, limiting the amount of observable information available for forensic inference. Finally, Zigbee commonly relies on a mesh network topology, in which devices forward traffic on behalf of others rather than communicating directly with a central access point, as is the case with Wi-Fi. This architectural feature further complicates traffic analysis, as relay operations introduce noise into traffic traces and such noise is inherently time-varying due to dynamic routing path adaptations. As a result, forensic insights derived from traffic analysis in a given Zigbee network topology may not generalize to other topologies, in contrast to what is more commonly observed in Wi-Fi–based environments.
This study addresses the aforementioned challenges through the following main contributions:
\begin{itemize}
    \item \textit{ZIOTP2025 traffic dataset\footnote{https://github.com/antonio-boiano/ZIOTP2025}:} 
    We collected and publicly released a comprehensive dataset of Zigbee network traffic captured from 21 off-the-shelf smart home devices deployed under two distinct network configurations. The dataset includes both background traffic and traffic generated through controlled user interactions, and is fully labeled to support IoT forensic tasks such as device identification, event detection, and traffic classification. To the best of the authors’ knowledge, this represents the largest publicly available labeled Zigbee traffic dataset specifically designed for IoT forensics.

    \item \textit{Traffic analysis:} 
    Leveraging a \gls*{soa}-based methodology, we developed a \gls*{ml} framework for Zigbee device classification, addressing two core tasks: device type classification and individual device identification. To mitigate the complexity introduced by multi-hop communications in mesh network topologies—where devices forward packets on behalf of others—we restrict the analysis to first-hop messages, thereby reducing bias due to retransmission and relaying behavior. Nevertheless, we show that temporal traffic features remain strongly affected by topology-dependent dynamics. Experimental results demonstrate that, while intra-topology classification achieves excellent performance (with F1-scores above 0.95), cross-topology generalization suffers from significant degradation. This highlights the critical and often overlooked impact of Zigbee mesh topology on the generalizability of traffic-based classification models.

    \item \textit{Storage--accuracy analysis:} 
    We investigate the trade-off between traffic storage requirements and the accuracy of downstream forensic analysis tasks. Specifically, we compare lossless traffic compression—obtained by applying state-of-the-art compression algorithms directly to raw Zigbee traffic—with lossy compression of traffic representations, where scalar quantization is applied to the statistical features used as input to \gls*{ml} models. Our results show that the latter approach enables a substantial reduction in storage bitrate without degrading classification accuracy, making it a promising solution for IoT forensic systems tasked with long-term storage of large-scale traffic traces.
\end{itemize}


The remainder of this paper is organized as follows. 
Section \ref{background} introduces the Zigbee protocol architecture, while Section \ref{state-of-the-art} reviews related work in IoT forensics and Zigbee traffic analysis. 
Section \ref{ziotp2025} presents the ZIOTP2025 dataset and the adopted data collection methodology. 
Section \ref{sec:methodology} describes the feature extraction and classification framework, and Section \ref{results} discusses the experimental results under both intra-topology and cross-topology settings. 
Section \ref{sec:compression} analyzes the trade-off between traffic storage and classification accuracy. 
Finally, Section \ref{sec:conclusion} concludes the paper and outlines directions for future research.

%% file: Chapters/background_v2.tex
\section{The Zigbee Protocol}
\label{background}
Zigbee is a low-cost, low-power wireless communication protocol designed for short-range communications, making it well suited for embedded systems and smart home applications. Its support for mesh networking enables the deployment of robust and self-healing networks with extended coverage. Zigbee is built on top of the IEEE 802.15.4 standard \cite{ieee_802_15_4}, which specifies low-rate wireless personal area networks (LR-WPANs) optimized for low data rates and minimal energy consumption.

\subsection{Zigbee Network Architecture}
The Zigbee protocol stack consists of four layers: the physical (PHY) layer, the medium access control (MAC) layer, the network (NWK) layer, and the application (APL) layer. The PHY and MAC layers are defined by IEEE 802.15.4 and are responsible for physical transmission and channel access control.

IEEE 802.15.4 operates over multiple frequency bands to accommodate regional regulations. In the widely adopted 2.4~GHz band, it supports 16 channels with 5~MHz spacing. The MAC layer employs a contention-based access mechanism based on CSMA/CA. In beacon-enabled networks, the PAN Coordinator periodically transmits beacon frames containing network parameters and synchronization information.

Building on IEEE 802.15.4, Zigbee introduces the NWK and APL layers to enable multi-hop routing, network management, and application-level services. The application layer includes the Application Support Sublayer (APS), the Zigbee Device Object (ZDO), and an application framework defining device-specific operations. The APS handles data binding, fragmentation, and security, while the ZDO is responsible for network formation, device discovery, and security configuration. Although Zigbee application profiles promote interoperability, their standardized data formats can increase the complexity of traffic-based device classification.

Zigbee networks comprise three main device roles: the \gls*{panc}, Full Function Devices (FFDs), and Reduced Function Devices (RFDs). The \gls*{panc} manages network creation, addressing, and coordination. FFDs, commonly acting as routers, relay packets on behalf of other devices and enable multi-hop communication, which allows Zigbee networks to extend coverage without relying on a high-power central access point. RFDs are typically low-power sensors or actuators that communicate through routers or directly with the \gls*{panc}.

Each Zigbee device is uniquely identified by a 64-bit IEEE MAC address (extended address). After joining the network, a 16-bit short address is assigned by the coordinator and used for subsequent intra-network communications.

These architectural characteristics have direct implications for traffic observability and analysis, particularly in forensic scenarios involving multi-hop communications and dynamically evolving network topologies.


%% file: Chapters/sota_v2.tex
\section{State of the Art}
\label{state-of-the-art}
This section reviews prior work related to the main contributions of this study, covering the availability of Zigbee network traffic datasets, existing approaches to Zigbee traffic analysis and profiling, and prior efforts on the trade-off between traffic storage requirements and analysis accuracy, which have so far been explored primarily in Wi-Fi and IP-based network settings.

\subsection{Dataset Availability}
Recent survey works provide a comprehensive overview of publicly available network traffic datasets for IoT systems, consistently highlighting the severe scarcity of datasets containing realistic Zigbee traffic. In particular, out of 74 datasets surveyed in \cite{de2023survey}, only seven include Zigbee traffic, and only a small subset of these provide packet-level captures suitable for fine-grained traffic analysis, such as device fingerprinting or physical-layer studies.

Among the publicly available datasets, the work in \cite{c7-nvc6-4q28-22} is the only one that includes traffic from ten commercial Zigbee devices commonly deployed in smart home environments. Traffic was captured using a software-defined radio platform (USRP N210) with GNU Radio and the gr-ieee802-15-4 and gr-foo modules, enabling IEEE 802.15.4 packet reception and storage in PCAP format. However, the dataset is released exclusively as unlabeled raw traffic traces, limiting its applicability to supervised IoT forensic tasks.

Most other studies rely on proprietary or non-public datasets. For instance, Sadikin et al. \cite{sadikin2020zigbee} developed a Zigbee intrusion detection system and collected attack traces involving one gateway and three smart light bulbs, but did not release the dataset. Bihl et al. \cite{bihl2016feature} conducted physical-layer fingerprinting experiments on ten identical Texas Instruments CC2420 Zigbee devices by collecting RF emissions in controlled environments. Patel et al. \cite{patel2014improving} extended this line of work to 13 CC2420 devices and evaluated ensemble learning techniques, such as Random Forest and AdaBoost, achieving improved identification performance compared to parametric classifiers. Despite their promising results, the datasets used in these studies are not publicly available.

Among the few public datasets that include Zigbee traffic, Dadkhah et al. \cite{iot_dataset_paper} introduced a multidimensional IoT profiling dataset primarily focused on IP-based devices, with a limited Zigbee subset. This subset includes nine distinct Zigbee devices (16 instances) from four manufacturers, captured under controlled laboratory conditions. While noteworthy for incorporating non-IP IoT technologies, the Zigbee component of the dataset exhibits limited interaction diversity, as only single, non-repetitive device–action pairs were recorded. Such constraints limit its suitability for comprehensive traffic profiling in complex and distributed environments such as smart homes.
Lourme et al. \cite{NDW74U_2023} and Keleşoğlu et al. \cite{app142310844} released ZBDS2023 and ZigBeeNet, respectively: multi-day Zigbee traffic traces collected from 10 and 14 devices, with RSSI information available. ZBDS2023 further provides multi-point captures using multiple sniffers deployed at different locations within the home. Despite these strengths, both datasets offer limited support for evaluating generalization in realistic smart-home settings, as the devices are largely drawn from a single vendor ecosystem (predominantly Philips) and the captures do not span multiple network topologies, preventing robustness assessments of ML models across diverse deployment conditions.

Motivated by these limitations, the present work introduces a dedicated Zigbee data acquisition campaign aimed at increasing both interaction diversity and traffic volume. In contrast to prior datasets, our campaign captures traffic from 21 Zigbee devices spanning multiple functional classes, including temperature sensors, door sensors, motion detectors, smart plugs, smart bulbs, vibration sensors, and light sensors, and collects data in multi-topology environments where devices operate concurrently rather than in isolation. This setup enables a systematic analysis of how network topology and deployment conditions affect the performance and generalization of state-of-the-art traffic-profiling methodologies.


\subsection{Zigbee Network Traffic Analysis}
Despite the extensive literature on IoT traffic classification, relatively few studies focus specifically on Zigbee-based traffic analysis. Rong Li et al. \cite{ZPA2023} investigated potential privacy leakage from encrypted Zigbee traffic, aiming to infer device type and operational state. Their approach extends classical feature-based traffic analysis by incorporating inter-event timing information, along with selected protocol-level features such as MAC addresses and channel identifiers. Machine learning models were then employed for device classification, identification, and event detection.

Similarly, the authors of \cite{ziot2020} proposed a fingerprinting framework for the passive identification of device classes in Zigbee and Z-Wave networks. Their method primarily relies on inter-arrival time (IAT) statistics of network frames, achieving effective device-class identification. Other works, such as \cite{Peek-a-Boo2018,antonio}, also explored Zigbee traffic analysis, but were limited by the use of small-scale datasets.

Additional studies have examined encrypted Zigbee traffic from different perspectives. Some focus on exploiting protocol vulnerabilities to infer sensitive device information, such as logical device roles \cite{zigator20} or spatial location \cite{Localize2022}. Others have released Zigbee datasets targeting either device profiling \cite{iot_dataset_paper} or security applications, such as the dataset introduced in \cite{ZBDS2023} for intrusion detection. In parallel, several works have addressed Zigbee device fingerprinting using physical-layer characteristics rather than network traffic features \cite{27patel2015non,28bihl2016feature,30dubendorfer2013zigbee}.

While existing studies demonstrate the feasibility of Zigbee traffic classification, they generally overlook the impact of mesh network topology on model robustness. Unlike IP-based networks, Zigbee extensively relies on multi-hop communication, where changes in network topology may significantly affect observed traffic patterns. In this work, we build upon methodologies proposed in the state of the art to systematically quantify performance degradation across different network topologies, providing a baseline for machine-learning-based Zigbee traffic profiling in realistic large-scale smart home environments.

\subsection{Storage-accuracy analysis}
The trade-off between traffic storage requirements and the accuracy of downstream forensic analysis has received limited attention in IoT forensics, where most approaches assume the availability of complete traffic traces and overlook long-term storage scalability issues.
An initial study of this problem was presented in \cite{fabio2,fabio3}, which proposed a framework for optimizing storage efficiency while preserving analysis accuracy. Focusing on Wi-Fi and IP-based IoT traffic, the authors compared lossless compression of raw traces with lossy compression of extracted traffic features, showing that feature-level compression can substantially reduce storage requirements without degrading classification performance.
More recently, Palmese et al. \cite{palmese2025resource} investigated resource optimization for evidence collection and preservation in forensics-ready access points, analyzing how storage and processing constraints affect IP-based traffic monitoring at the network edge.
Despite these contributions, existing works remain limited to Wi-Fi and IP-based technologies and do not consider the distinctive characteristics of Zigbee networks, such as low traffic rates, small packet sizes, and mesh-based multi-hop communication. This work extends the storage--accuracy analysis to Zigbee traffic, evaluating compression strategies in realistic multi-topology smart home scenarios.

%% file: Chapters/dataset_v2.tex
\section{ZIOTP2025: A Dataset for Zigbee traffic classification}
\label{ziotp2025}
Despite the widespread adoption of Zigbee in smart home environments, publicly available Zigbee traffic datasets remain scarce. To address this gap and support reproducible research, we introduce ZIOTP2025, a dataset of Zigbee network traffic collected over multiple days in realistic smart home settings and capturing natural user-device interactions. The data collection methodology builds upon prior work \cite{iot_dataset_paper}, extending its Zigbee component by capturing traffic from 21 off-the-shelf devices spanning 15 models and seven manufacturers.
To enable the investigation of topology-dependent effects, devices were deployed across two distinct Zigbee personal area networks (PANs) during data acquisition. This multi-topology design allows systematic analysis of how network configuration impacts traffic-based forensic tasks such as device classification and identification.

\subsection{Acquisition Setup}
\label{Data_acquisition_Setup}
A dedicated network architecture was deployed to replicate a realistic smart home environment. An isolated \gls*{lan} was configured using a consumer-grade router running OpenWRT, with a Raspberry Pi running Home Assistant OS \cite{home_assistant_os} acting as the central smart home gateway and Zigbee \gls*{panc}. Zigbee connectivity was provided through a USB radio interface managed by the \gls*{zha} integration\footnote{https://www.home-assistant.io/integrations/zha/}, which handled network formation and device onboarding. For all experiments, the \gls*{pan} was configured on Channel~11 and devices were added in isolation to ensure reproducible initialization.

\begin{figure}[]
    \centering
    \includegraphics[width=\columnwidth]{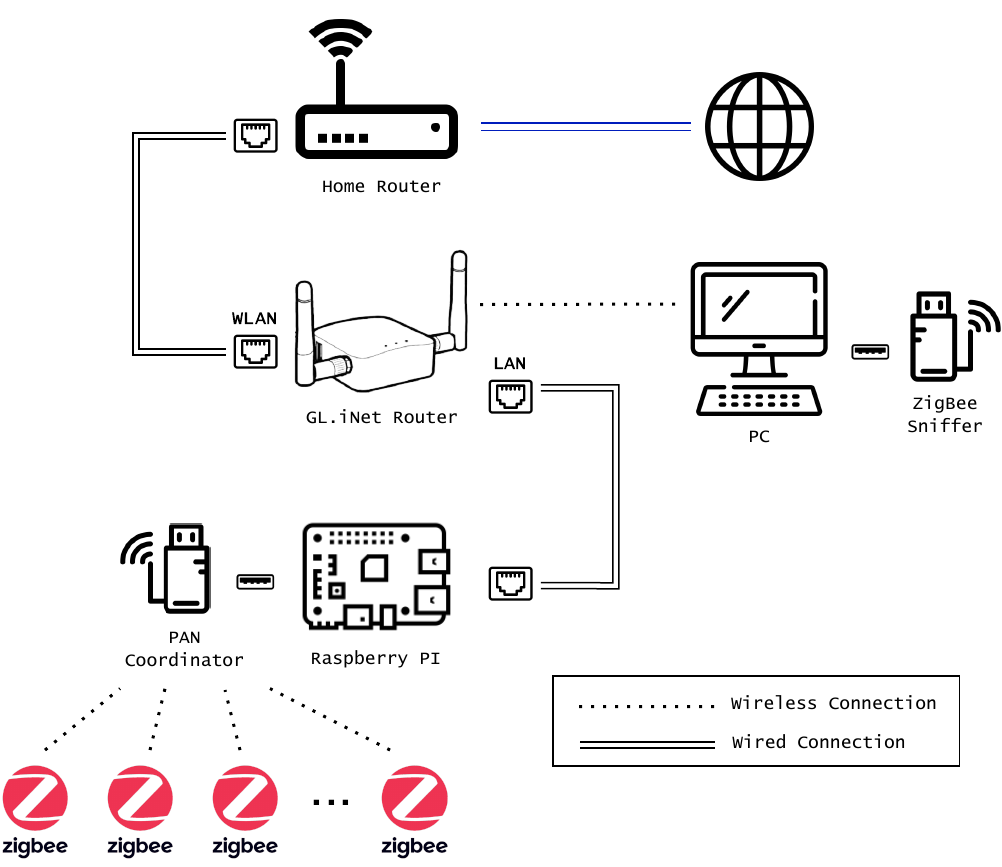}
    \caption{Network configuration of the smart home and acquisition pipeline used to capture Zigbee and IP traffic from commercial IoT devices. The smart home is managed by a central gateway acting as PAN coordinator, while a dedicated sniffer and a monitoring host capture Zigbee and IP network traces within an isolated network.}
    \label{fig:Chapter4:network_configuration}
\end{figure}

IP-based traffic exchanged within the \gls*{lan} was captured from a host connected to the network, which also provided access to the Home Assistant dashboard. Zigbee traffic at the IEEE~802.15.4 layer was collected using a dedicated packet sniffer operating in promiscuous mode. To ensure reliable capture, the sniffer and the Zigbee radio interface were placed in close proximity. The overall network configuration is illustrated in Figure~\ref{fig:Chapter4:network_configuration}.

\begin{figure}
\centering
\subfigure[Topology A]{ \includegraphics[width=0.4\textwidth,trim={10 14 10 15},clip]{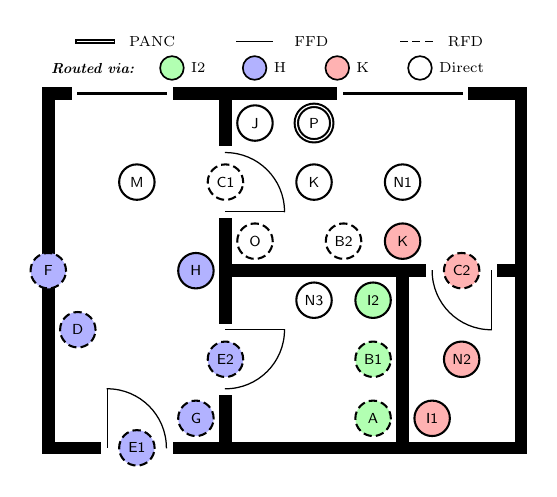} \label{fig:subfig1_A}} 
\subfigure[Topology B]{ \includegraphics[width=0.39\textwidth,trim={10 14 10 14},clip]{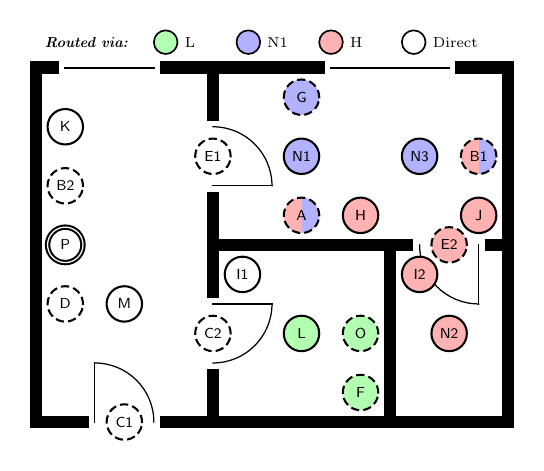} \label{fig:subfig2_B}} 
\caption{Device disposition during the acquisition campaign for the two considered topologies.
Devices are positioned according to their physical deployment and labeled as reported in Table \ref{tab:dataset}.
The outer line style identifies the device role (PANC, FFD, or RFD).
Color coding indicates the logical communication path toward the PANC: white denotes devices directly associated with the PANC, while colored devices communicate with the PANC via an intermediate FFD acting as a router.} 
\label{fig:Chapter4:device_topology} 
\end{figure}

To analyze the impact of network topology on Zigbee traffic analysis, data acquisition was conducted under two distinct Zigbee network topologies. Here, topology refers to the spatial arrangement of devices and the resulting communication paths among the \gls*{panc}, \glspl*{ffd}, and \glspl*{redfd}. Unlike Wi-Fi’s star topology, Zigbee relies on mesh communication, where messages may be forwarded through intermediate \glspl*{ffd}. The two adopted topologies, shown in Figure~\ref{fig:Chapter4:device_topology}, differ in device placement and coordinator position, inducing distinct multi-hop communication patterns. The acquisition procedure was kept identical across both topologies.

\subsection{Acquiring Device Behaviors}
\label{acq_behav}
Following the acquisition methodology introduced in \cite{iot_dataset_paper}, a set of controlled interaction scenarios was designed to capture a wide range of Zigbee device behaviors and operational states. The acquisition campaign included: (i) idle periods with no user interaction, used to characterize background and periodic traffic; (ii) power cycling events for mains-powered devices, capturing network join and leave procedures; (iii) user-issued commands executed through the smart home controller; (iv) direct physical interactions with sensors (e.g., motion, door, vibration); and (v) extended active scenarios emulating natural day-to-day smart home usage. Each scenario was repeated multiple times to capture temporal variability and enrich the dataset with diverse traffic patterns.

\subsection{Ground Truth Extraction}
Since the dataset captures encrypted Zigbee traffic, device identity and application-layer payloads are not directly observable. Device identities were recovered a posteriori by isolating each device and associating its traffic with the corresponding IEEE MAC address. To enable event-level labeling, each device was individually reset and paired with the \gls*{panc}, while Zigbee traffic exchanged during the pairing phase was captured.

During commissioning, the transport key is exchanged within the Zigbee Application Support (APS) layer and encrypted using a publicly known trust center link key. Leveraging this mechanism, pairing traffic was decrypted using Wireshark\footnote{https://www.wireshark.org/} to extract the device-specific transport key, which was then used exclusively for dataset labeling and validation.

All subsequent traffic analysis and machine learning experiments were performed on encrypted traffic only, without relying on decrypted payload information.

\subsection{Dataset Characteristics and Availability}
The ZIOTP2025 dataset comprises 252,896 IEEE~802.15.4 frames, including 69,497 Zigbee frames corresponding to approximately 2,520 human-generated commands collected across two network topologies. The traffic volume is nearly balanced between the two configurations, with Topology~B containing approximately 3\% more data than Topology~A. Each topology-specific dataset occupies approximately 17~MB and includes traffic generated by the \gls*{panc}, mains-powered devices (e.g., smart bulbs and plugs), and battery-operated devices. In addition, around 40~MB of IP-based traffic captured during command interaction phases is provided, enabling cross-technology analysis.
The dataset includes traffic from 21 commercial Zigbee devices spanning multiple categories (e.g., sensors, actuators, and lighting devices) and several manufacturers. Multiple instances of selected devices are included to support comparative analysis across devices of the same model. Table~\ref{tab:dataset} summarizes device models, quantities, and per-device traffic contributions. As expected, traffic volume is highly unbalanced across device roles, with \glspl*{redfd} generating up to two orders of magnitude fewer packets than \glspl*{ffd}. This imbalance reflects the low-duty-cycle design of battery-powered devices and introduces additional challenges for machine-learning-based traffic analysis.
The ZIOTP2025 dataset is publicly available to support reproducible research in Zigbee traffic analysis and IoT forensics.

\begin{table*}[]
\centering
\begin{tabular}{c c c c c c c c}
\textbf{Label} &\textbf{Brand} & \textbf{Device} & \textbf{Category} & \textbf{Model} & \textbf{Qty} & \textbf{Type} & \textbf{Data Dist.} \\\hline\\[-5px]
A &Sonoff & Temperature Sensor& Temperature & SNZB-02 & 1 & RFD &0.25\% \\[2px]
B &Sonoff & Motion Sensor& Motion& SNZB-03 & 2  & RFD &0.12\%\\[2px]
C &Sonoff & Door Sensor& Door& SNZB-04 & 2 & RFD  &0.10\%\\[2px]
D &Aqara & Motion Sensor P1 & Motion& MS-S02 & 1 & RFD &0.22\% \\[2px]
E &Aqara & Door and Window Sensor & Door& DW-S02D & 2 & RFD & 0.32\% \\[2px]
F &Aqara & Vibration Sensor & Motion& DJT11LM & 1 & RFD &0.17\%\\[2px]
G &Aqara & Mini Switch & Button& WB-R02D & 1 & RFD &0.33\%\\[2px]
H &Tuya & Smart Socket EU-16A & Socket& N/A & 1  & FFD &16.44\% \\[2px]
I &Senckit & Smart Power Plug & Socket&  N/A & 2 & FFD &15.38\% \\[2px]
J &Ledvance & Z3 Plug & Socket& 4058075208513 & 1 & FFD &3.08\% \\[2px]
K &Ledvance & Smart+ Plug & Socket&  4058075729261 & 1 & FFD &2.06\% \\[2px]
L &Ledvance & Smart+ Bulb &  Bulb& 4058075208391 & 1 & FFD  &8.77\% \\[2px]
M &Moes & Bulb &  Bulb&  GU10 & 1 & FFD &4.12\% \\[2px]
N &Philips & HueLamp & Bulb& LWA019 & 3 & FFD &5.75\%  \\[2px]
O &Philips & HueMotion & Motion& SML001 & 1 & RFD &4.24\%  \\[2px]
\end{tabular}
\\[5pt]
\caption{List of Zigbee IoT devices included in the dataset, with information on: brand, device type, model identifier, and quantity. Device identifiers are assigned systematically using alphabetical labels, with numerical suffixes appended when multiple units of the same device type are present. The table additionally presents the per-device data distribution, quantifying each device's proportional contribution to the total volume of Zigbee packets captured in the two topologies.}
\label{tab:dataset}
\end{table*}

%% file: Chapters/methodology_v3.tex
\section{Traffic Analysis Methodology}
\label{sec:methodology}
This section describes the methodology adopted to analyze encrypted Zigbee traffic and evaluate traffic-based device profiling under different network topologies. Two classification tasks are considered: \textit{device type classification} and \textit{ndividual device identification}. The former targets coarse-grained categorization of devices based on functional roles and communication patterns, while the latter addresses fine-grained discrimination among individual devices. Although Zigbee enforces standardized message formats, device-specific characteristics remain observable due to differences in implementation, hardware, and operational behavior.

\subsection{Data Preprocessing}
\label{sec:data_pre}
Raw Zigbee traffic captured in PCAP format was processed through a preprocessing pipeline designed to generate structured representations suitable for machine-learning-based analysis.

\paragraph{Ground truth and metadata extraction}
Each packet was associated with device-level metadata by mapping 16-bit Zigbee short addresses to device categories and labels (Table~\ref{tab:dataset}). Payload decryption was used exclusively during dataset labeling to distinguish human-triggered interactions from background traffic, while all subsequent analysis relied solely on encrypted traffic. Packet-level attributes extracted include source and destination identifiers at the IEEE~802.15.4 and Zigbee NWK layers, packet length, and timestamp.

\paragraph{Filtering and feature extraction}
To capture temporal dependencies, packet traces were segmented into non-overlapping time windows, each aggregating consecutive packets. For each device, bidirectional traffic exchanged with the \gls*{panc} was isolated, and relayed packets introduced by mesh routing were filtered by comparing Zigbee-layer source addresses with IEEE~802.15.4 transmitter addresses.

For each window containing at least two packets, statistical features were extracted from packet lengths and inter-arrival times, computed separately for uplink, downlink, and aggregated traffic. Extracted metrics include mean, standard deviation, minimum, maximum, median, skewness, kurtosis, mean absolute deviation, and packet count.

\subsection{Classification and Validation Strategy}
\label{subsec:classifier}
Both classification tasks were addressed using eXtreme Gradient Boosting (XGBoost), which is well suited for multi-class classification on structured feature sets. Hyperparameters were optimized independently for each task using grid search with cross-validation.

For device type classification, the model employed 300 estimators with a maximum tree depth of 8 and moderate regularization. Individual device identification required a more expressive configuration, using the same number of estimators but deeper trees (maximum depth of 10) and additional instance subsampling to improve robustness. All remaining parameters were left at default values.

Model evaluation was performed using stratified 5-fold cross-validation to preserve class distributions. Performance was assessed using both weighted and macro-averaged F1-scores. Unless otherwise stated, results are reported using the macro F1-score, which is particularly sensitive to performance on minority classes and therefore suitable for the imbalanced nature of the dataset.

%% file: Chapters/results_v3.tex
\section{Performance Evaluation}
\label{results}
This section evaluates the performance of the traffic analysis methodology introduced in Section~\ref{sec:methodology}. We assess the effectiveness of the XGBoost-based classifier for the two traffic analysis tasks considered in this work—device type classification and individual device identification—under both intra-topology and cross-topology evaluation settings.

The evaluation is structured as follows. First, we perform a preliminary analysis to identify the optimal temporal window size for feature extraction. We then report classification performance when training and testing are conducted within the same network topology, representing the standard evaluation scenario adopted in the literature. Finally, we analyze the generalization capability of the model by evaluating performance across different network topologies.

\subsection{Optimal Time Window Selection}

\begin{figure}[htbp]
\centering
\subfigure[Device Type Classification]{ 
    \includegraphics[width=0.504\linewidth,trim={8 8 7 7},clip]{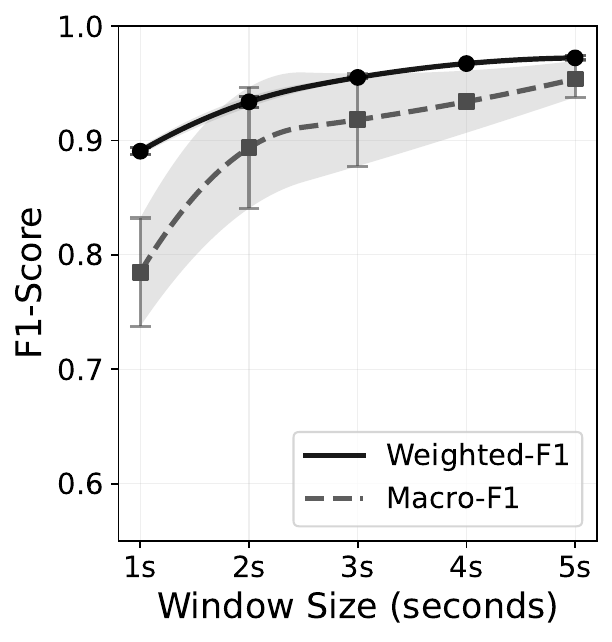} 
    \label{fig:device_type_classification}
} 
\subfigure[Individual Device Identification]{ 
    \includegraphics[width=0.425\linewidth,trim={8 8 6 7},clip]{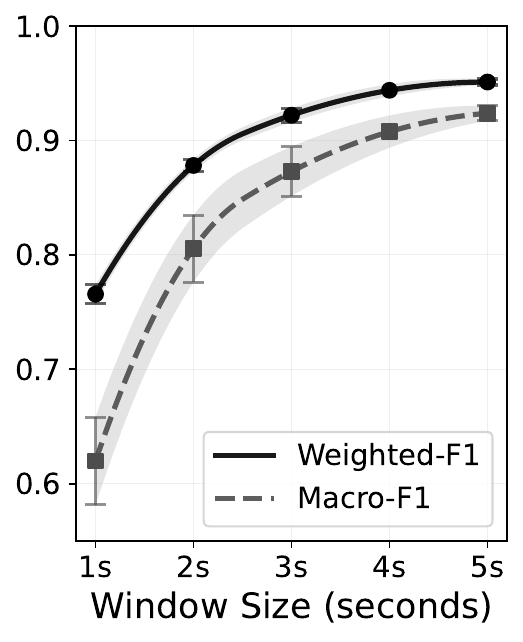} 
    \label{fig:device_name_classification}
} 
\caption{Weighted and Macro F1-Score across different window sizes for: (a) device type classification and (b) individual device identification classification task.} 
\label{fig:f1_score_comparison} 
\end{figure}

Before proceeding with the main classification experiments, we conducted a preliminary study to determine the optimal temporal window size for feature extraction. Since our approach aggregates network traffic data using time windows, this parameter critically affects the trade-off between capturing sufficient temporal context and maintaining adequate sample sizes for robust classification.
We tested five different window lengths (1, 2, 3, 4, and 5 seconds), training and evaluating the XGBoost classifier for each configuration. Figure \ref{fig:f1_score_comparison} presents the weighted and macro F1-scores across these window sizes, revealing a consistent trend for both tasks: classification performance improves steadily with increasing window size, as longer windows more effectively capture meaningful temporal patterns in network traffic data.
Specifically, for device type classification, the weighted F1-score increased from 0.89 (1-second window) to 0.97 (5-second window), and the macro F1-score improved from 0.78 to 0.95. For individual device identification, in the same range of window sizes, the weighted F1-score increased from 0.76 to 0.95, and the macro F1-score from 0.62 to 0.92.
Both tasks exhibit performance curves approximating a saturating exponential function, characterized by rapid initial gains followed by progressively diminishing returns. The marginal improvement beyond the 4-second threshold is minimal, suggesting that a 5-second window captures nearly all relevant temporal dynamics. Based on this empirical characterization and considering the trade-off between classification performance, computational efficiency, and prediction time, we fixed the sliding window size at 5 seconds for all subsequent experiments presented in the next sections.

Before presenting the classification results, a statistical analysis of the ML-ready dataset (described in Section \ref{sec:data_pre}) revealed that \gls*{redfd} devices, which typically transmit single Zigbee packets in response to specific events, can produce windows containing insufficient packets (i.e., windows with fewer than 2 packets) for meaningful feature extraction. Consequently, two devices: the Sonoff temperature sensor and the Aquara Vibration sensor, yielded fewer than 10 valid 5-second windows across both topologies and were excluded from further analysis.

Based on these results, a 5-second temporal window was selected for all subsequent experiments.

\subsection{Intra-topology Classification}
We first evaluate classification performance under an intra-topology setting, where training and testing data originate from the same Zigbee network topology. This scenario represents the standard evaluation setup adopted in prior work on IoT traffic classification and serves as a baseline for subsequent cross-topology analysis. All results are obtained using stratified 5-fold cross-validation.

\subsubsection{Device Type Classification}
\begin{figure}[htbp]
\centering
\includegraphics[width=\linewidth]{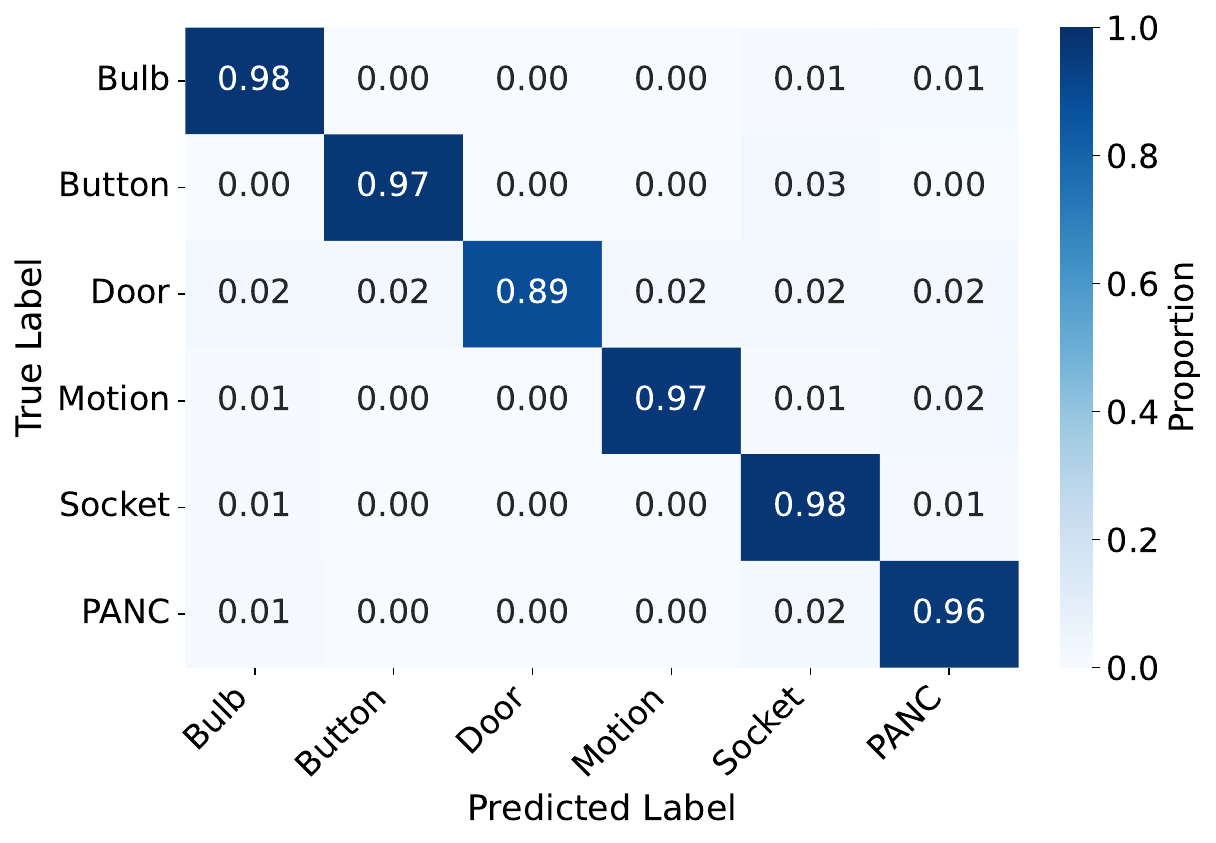}
\caption{Normalized Confusion matrix for device type classification on a 5-second window in topology B.}
\label{fig:cls-type_topo_B}
\end{figure}

The proposed classifier achieves excellent performance when evaluated under the intra-topology setting. Using a 5-second temporal window and topology~B, the model attains a weighted F1-score of 0.97 and a macro F1-score of 0.95. The corresponding classification report and normalized confusion matrix are shown in Table~\ref{tab:cls-type_topo_B} and Figure~\ref{fig:cls-type_topo_B}, respectively.

High precision and recall are consistently observed across nearly all device categories, indicating strong discriminative capability between different smart home device types. Comparable results are obtained for topology~A, as summarized in Table~\ref{tab:classification_results_type}.

\begin{table}[htbp]
    \centering
    \begin{tabular}{lccc}
    \toprule
    \textbf{Class} & \textbf{Precision} & \textbf{Recall} & \textbf{F1-score} \\
    \midrule
    Bulb         & 0.94 & 0.98 & 0.96  \\
    Button       & 0.92 & 0.97 & 0.95 \\
    Coordinator  & 0.98 & 0.96 & 0.97 \\
    Door         & 0.98 & 0.89 & 0.93 \\
    Motion       & 0.92 & 0.97 & 0.94 \\
    Socket       & 0.98 & 0.98 & 0.98 \\
    \midrule
    \textbf{Accuracy}     &       &       & 0.97 \\
    \textbf{Macro Avg}    & 0.95 & 0.96 & 0.95  \\
    \textbf{Weighted Avg} & 0.97 & 0.97 & 0.97 \\
    \bottomrule
    \end{tabular}
    \caption{Classification report for device type classification on a 5-second window in topology B.}
    \label{tab:cls-type_topo_B}
\end{table}

\subsubsection{Individual Device Identification}
Despite the increased complexity of individual device identification, the classifier maintains strong performance under the intra-topology setting. With 14 distinct device identities, the model achieves a weighted F1-score of 0.95 and a macro F1-score of 0.93 when evaluated on topology~B.

\begin{figure*}[htbp]
\centering
\includegraphics[width=0.9\linewidth]{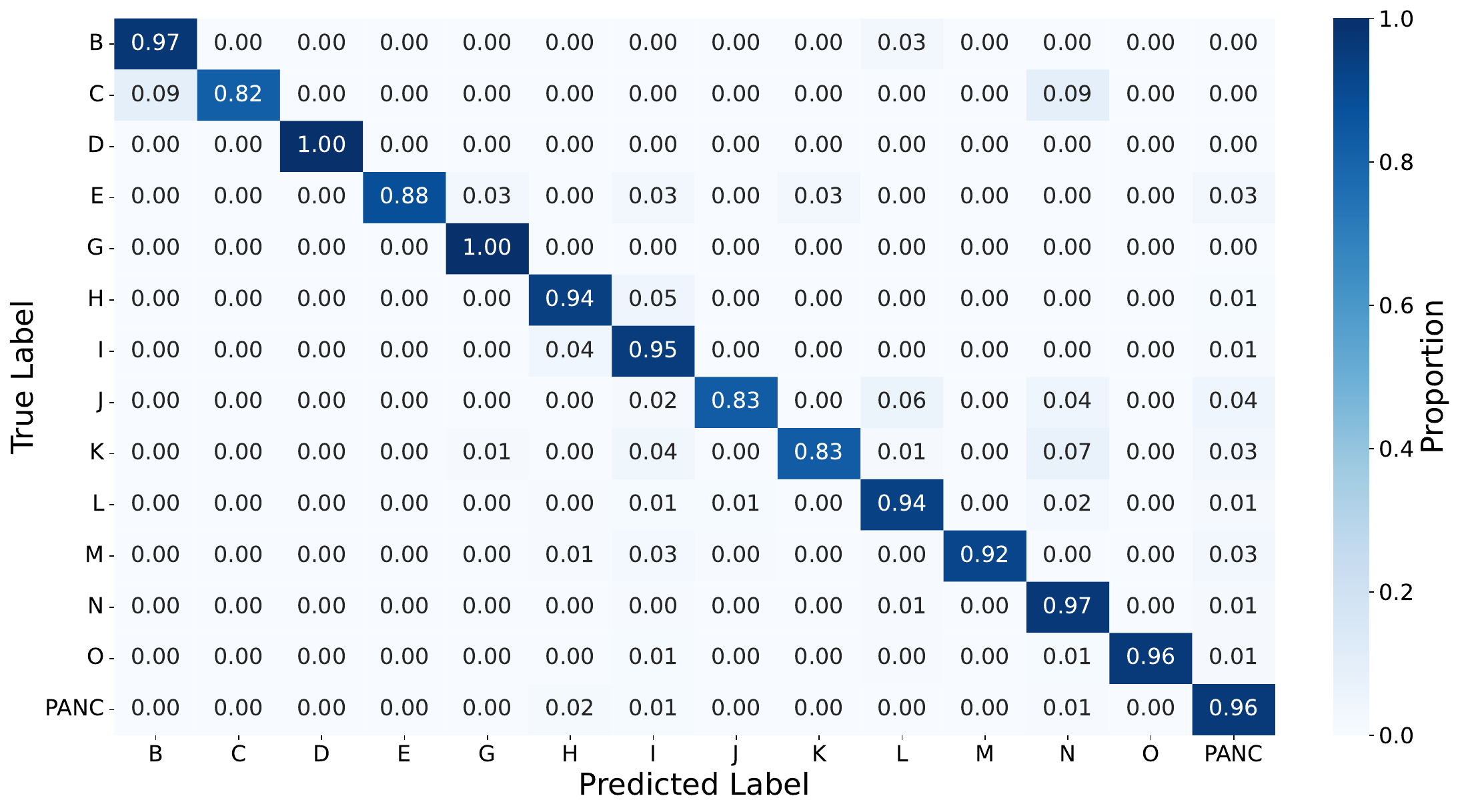}
\caption{Normalized Confusion matrix for individual device identification on a 5-second window in topology B.}
\label{fig:cls-name_topo_B}
\end{figure*}

The normalized confusion matrix and classification report, shown in Figure~\ref{fig:cls-name_topo_B} and Table~\ref{tab:device_name_classification_topoB}, respectively, indicate robust performance across most devices. The \gls*{panc} is particularly well distinguished, likely due to its unique role in network coordination. Results for both topology~A and B are reported in Table~\ref{tab:classification_results_name}.

\begin{table}[htbp]
    \centering
    \begin{tabular}{llccc}
    \toprule
    \textbf{Class} & \textbf{Class Name} & \textbf{Precision} & \textbf{Recall} & \textbf{F1-score} \\
    \midrule
  B & Sonoff Motion Sensor  & 0.84 & 0.97 & 0.90 \\
  C & Sonoff Door Sensor & 0.90 & 0.82 & 0.86 \\
  D & Aqara Motion Sensor & 1.00 & 1.00 & 1.00 \\
  E & Aqara Door/Window Sensor & 1.00 & 0.88 & 0.94 \\
  G & Aqara Mini Switch & 0.92 & 1.00 & 0.96 \\
  H & Tuya Smart Socket & 0.91 & 0.94 & 0.92 \\
  I & Senckit Smart Power Plug & 0.95 & 0.95 & 0.95 \\
  J & Ledvance Z3 Plug & 0.87 & 0.83 & 0.85 \\
  K & Ledvance Smart+ Plug  & 0.88 & 0.83 & 0.85 \\
  L & Ledvance Smart+ Bulb  & 0.94 & 0.94 & 0.94 \\
  M & Moes Bulb & 0.95 & 0.92 & 0.94 \\
  N & Philips HueLamp  & 0.93 & 0.97 & 0.95 \\
  O & Philips HueMotion  & 0.92 & 0.96 & 0.94 \\ 
  PANC &Pan Coordinator & 0.98 & 0.96 & 0.97 \\
    \midrule
    \textbf{Accuracy}     &&       &       & 0.95 \\
    \textbf{Macro Avg}    && 0.93 & 0.93 & 0.93  \\
    \textbf{Weighted Avg} && 0.95 & 0.95 & 0.95 \\
    \bottomrule
    \end{tabular}
    \caption{Classification report for individual device identification using a 5-second temporal window in topology B.}
    \label{tab:device_name_classification_topoB}
\end{table}

\begin{table*}[]
\centering
\caption{Classification Results Summary for Device Type classification task}
\label{tab:classification_results_type}
\begin{tabular}{|l|l|c|c|c|c|c|}
\hline
\textbf{Training} & \textbf{Test} &  \textbf{Accuracy} & \textbf{Precision} & \textbf{Recall} & \textbf{F1-Macro} & \textbf{F1-Weighted}  \\
\textbf{Dataset} & \textbf{Dataset} & & & & & \\
\hline
\multicolumn{7}{|c|}{\textbf{Intra-Topology (K-Fold, k=5)}} \\
\hline
 Topology A & Topology A  & 0.970 ± 0.003  & 0.949 ± 0.013 & 0.944 ± 0.022 & 0.945 ± 0.010 & 0.971 ± 0.003 \\
\hline
 Topology B & Topology B  & 0.972 ± 0.002  & 0.953 ± 0.018 & 0.957 ± 0.013 & 0.954 ± 0.016 & 0.972 ± 0.002 \\
\hline
\multicolumn{7}{|c|}{\textbf{Cross-Topology (Full Dataset)}} \\
\hline
Topology A & Topology B  & 0.878 & 0.738 & 0.765 & 0.745 & 0.879 \\
\hline
 Topology B & Topology A  & 0.850 & 0.724 & 0.759 & 0.741 & 0.849 \\
\hline
\end{tabular}
\end{table*}

\begin{table*}[]
\centering
\caption{Classification Results Summary for individual device identification task}
\label{tab:classification_results_name}
\begin{tabular}{|l|l|c|c|c|c|c|}
\hline
\textbf{Training} & \textbf{Test} &  \textbf{Accuracy} & \textbf{Precision} & \textbf{Recall} & \textbf{F1-Macro} & \textbf{F1-Weighted}  \\
\textbf{Dataset} & \textbf{Dataset} & & & & & \\
\hline
\multicolumn{7}{|c|}{\textbf{Intra-Topology (K-Fold, k=5)}} \\
\hline
  Topology A & Topology A  & 0.944 ± 0.004  & 0.910 ± 0.032 & 0.877 ± 0.017 & 0.889 ± 0.021 & 0.944 ± 0.004 \\
\hline
  Topology B & Topology B  & 0.951 ± 0.003  & 0.932 ± 0.016 & 0.924 ± 0.019 & 0.923 ± 0.007 & 0.951 ± 0.003 \\
\hline
\multicolumn{7}{|c|}{\textbf{Cross-Topology (Full Dataset)}} \\
\hline
Topology A & Topology B & 0.692 & 0.517 & 0.480 & 0.474 & 0.694 \\
\hline
 Topology B & Topology A  & 0.678 & 0.446 & 0.434 & 0.413 & 0.672 \\
\hline
\end{tabular}
\end{table*}

\subsection{Cross-topology Classification}
We next evaluate the generalization capability of the proposed traffic analysis methodology by adopting a cross-topology setting, in which training and testing are performed on data collected from different Zigbee network topologies. This evaluation exploits the multi-topology nature of the dataset and allows us to quantify the impact of topology-dependent traffic dynamics on classification performance.

\subsubsection{Device Type Classification}
Device type classification exhibits moderate performance degradation under cross-topology evaluation. When trained on one topology and tested on the other, the model achieves a weighted F1-score of approximately 0.85 and a macro F1-score of 0.75 (Table~\ref{tab:classification_results_type}).

While the \gls*{panc} remains consistently well classified, device categories dominated by \glspl*{redfd} experience larger performance drops. This behavior is attributable to limited training support and to topology-dependent routing effects introduced by the Zigbee mesh network. Overall, the classifier remains effective at distinguishing device types, reflecting the standardized command sets and communication patterns enforced by the Zigbee protocol.

\begin{figure}
    \centering
    \includegraphics[width=1\linewidth]{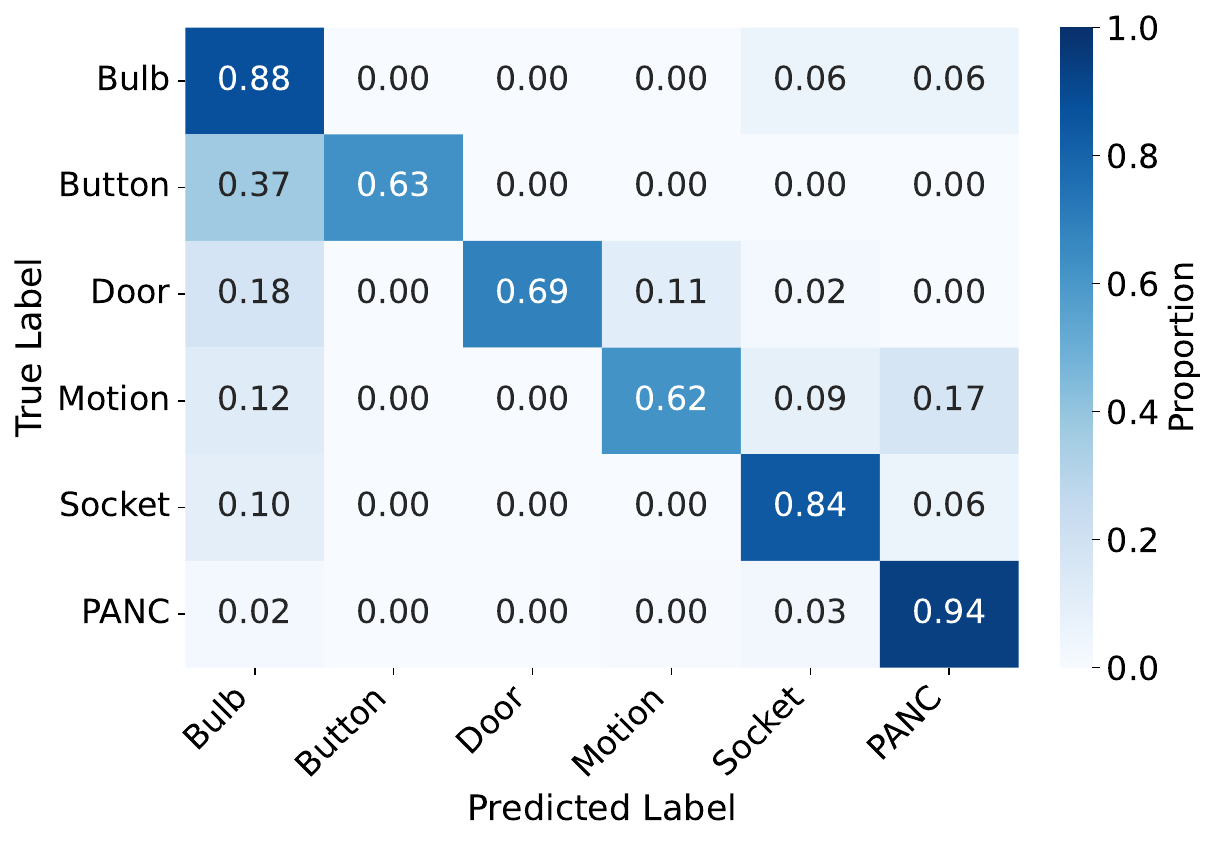}
    \caption{Normalized Confusion matrix for device type classification on cross-topology settings (trained on network traces acquired from topology B and tested against network traces coming from topology A.}
    \label{fig:cross-topo-cm}
\end{figure}
\subsubsection{Individual Device Identification}
In contrast, individual device identification is strongly affected by topology changes. When trained on topology~B and tested on topology~A, accuracy drops to 0.68, with a weighted F1-score of 0.67 and a macro F1-score of 0.41 (Table~\ref{tab:classification_results_name}).

The most significant degradation is observed for \glspl*{redfd}, whose average per-device F1-score falls below 0.2. This effect is exacerbated by the fine-grained nature of the task, which increases the number of classes and reduces the number of training samples per class. \glspl*{ffd} demonstrate comparatively higher robustness, with an average F1-score of 0.53, while the \gls*{panc} maintains stable performance (F1-score of 0.86), indicating largely topology-invariant traffic patterns.

Misclassifications predominantly occur among devices of the same manufacturer or device category, suggesting that manufacturer-specific traffic characteristics exhibit partial resilience to topology changes, whereas device-specific fingerprints are more sensitive to mesh routing dynamics.

\subsection{Discussion}
The experimental results highlight the effectiveness and the limitations of traffic-based device profiling in Zigbee networks. Under intra-topology conditions, the proposed methodology achieves excellent performance for both device type classification and individual device identification, confirming that encrypted Zigbee traffic retains sufficient behavioral information to support fine-grained analysis.

However, the cross-topology evaluation reveals a markedly different behavior between the two tasks. While device type classification shows only moderate performance degradation, individual device identification is significantly affected by changes in network topology. This divergence can be attributed to the standardized nature of Zigbee application profiles, which enforces consistent communication patterns at the device category level, in contrast to device-specific characteristics that are more sensitive to mesh routing dynamics and multi-hop communication paths.

The observed degradation is particularly pronounced for battery-powered \glspl*{redfd}, whose sparse and event-driven traffic, combined with relay operations performed by \glspl*{ffd}, reduces the stability of temporal traffic features across topologies. Conversely, mains-powered devices and the \gls*{panc} exhibit more robust and topology-invariant traffic patterns, resulting in higher classification resilience.

%% file: Chapters/compression_v3.tex
\section{Storage--Accuracy Analysis}
\label{sec:compression}
Captured Zigbee network traffic must often be retained to support a-posteriori IoT forensic investigations and long-term monitoring. However, the limited storage resources typically available on edge IoT controllers, combined with the continuous operation of smart home devices, make efficient traffic storage a critical design requirement.

In this section, we analyze the trade-off between storage efficiency and classification accuracy for Zigbee traffic analysis. We compare lossless compression techniques applied directly to raw packet traces with lossy approaches based on compact representations of traffic features. The goal is to quantify how different storage strategies impact the accuracy of downstream traffic classification tasks, and to identify solutions that enable substantial storage reduction without degrading forensic analysis performance.

\subsection{Lossless Compression of Raw Zigbee Packets}
We first evaluate the feasibility of long-term storage of captured Zigbee traffic by retaining raw packet traces in \texttt{.pcap} format and applying standard lossless compression techniques. This approach represents the most straightforward solution for traffic preservation, as it preserves full packet-level information and does not require any modification to the original data representation. At the same time, it establishes an upper bound on storage efficiency achievable without sacrificing information content.
Table~\ref{tab:lossless_compression} reports the storage footprint and average bitrate required to store raw Zigbee traffic captured under the two considered network topologies, together with the compression performance achieved using three widely adopted lossless compressors: GZIP, BZIP2, and 7Z-LZMA. These tools were selected due to their protocol-agnostic nature and widespread use in network trace archival.

Without compression, raw Zigbee traffic requires average storage rates of approximately 2.2~kbps and 2.9~kbps for Topology~A and Topology~B, respectively. Here, the storage rate is computed as the total size of the stored \texttt{.pcap} file divided by the corresponding capture duration, yielding an average bitrate over time. These values reflect the cumulative effect of long capture durations and continuous background traffic, despite the low data rate and small packet size characteristic of Zigbee communications.

Applying lossless compression substantially reduces storage requirements. GZIP and BZIP2 achieve comparable performance, reducing the bitrate by approximately 64--65\% across both topologies. Among the evaluated methods, 7Z-LZMA consistently provides the best compression performance, lowering the required storage rate to 657~bps for Topology~A and 768~bps for Topology~B, corresponding to an approximate 3.3× compression ratio relative to uncompressed \texttt{.pcap} files.

While lossless compression substantially reduces the storage footprint of raw Zigbee traffic, the resulting storage rates can still pose practical constraints in edge-based deployments characterized by limited storage resources and long-term evidence retention requirements. This motivates the exploration of alternative storage strategies based on compact traffic representations and controlled information loss.


\begin{table}[h]
    \centering
\caption{Evaluation of Lossless Compression Methods on the ZIOTP2025 PCAP Dataset: Size and Bitrate Across Two Network Topologies}
    \label{tab:lossless_compression}
    \begin{tabular}{lcccc}
    & \multicolumn{2}{c}{Topology A} & \multicolumn{2}{c}{Topology B} \\ \hline
Method & \multicolumn{1}{c}{Size (MB)} & \multicolumn{1}{c}{bps} & \multicolumn{1}{c}{Size (MB)} & \multicolumn{1}{c}{bps} \\
        \midrule
Raw & 6.69 & 2180 & 11.06 & 2880 \\
GZIP & 2.41 & 786 & 3.60 & 937 \\
BZIP2 & 2.36 & 771 & 3.51 & 913 \\
7Z-LZMA & 2.02 & 657 & 2.95 & 768
    \end{tabular}
\end{table}

\subsection{Storage of Traffic Features}
An alternative to storing raw packet traces consists in retaining compact representations of traffic behavior in the form of extracted features. This approach is particularly attractive when the primary objective is traffic analysis rather than packet-level reconstruction, as it eliminates the need for repeated feature extraction during offline processing.

However, unlike IP-based networks, Zigbee environments are characterized by low traffic intensity and small packet sizes. As a consequence, storing raw Zigbee packet traces—especially when combined with lossless compression—may already be relatively storage-efficient. In this setting, feature-based representations are not intrinsically more compact and may require higher storage rates when features are stored using standard high-precision formats.

To clarify this aspect, we consider the window-based feature extraction process described in Section~\ref{sec:data_pre}. Let $N$ denote the number of devices being monitored, $F$ the number of extracted features per window, $b$ the bit width used to encode each feature, and $T$ the window duration. Under these assumptions, the resulting average storage rate can be expressed as
\begin{equation}
    \frac{N \times F \times b}{T} \quad \text{bit/s}.
\end{equation}
This relationship highlights that, in Zigbee-based systems, the storage cost of feature-level representations can easily exceed that of raw packet traces when high-precision feature encodings are adopted.

These considerations indicate that feature-level storage does not automatically yield storage savings in Zigbee networks. Instead, any potential benefit must arise from reducing the precision of the feature representation. To this end, we investigate lossy compression of traffic features through scalar quantization, where each feature is encoded using a reduced number of bits.

Specifically, we generate quantized feature representations using bit depths ranging from 1 to 64 bits per feature. The resulting quantized feature vectors are subsequently compressed using a lossless encoder (7Z-LZMA) to remove residual redundancy. We evaluate two classification scenarios: (i) models trained on unquantized features and tested on quantized features, and (ii) models trained and tested on quantized features using the same bit depth. This analysis allows us to quantify the trade-off between feature precision, storage efficiency, and classification accuracy.

\begin{figure}
    \centering
    \subfigure[]{
        \includegraphics[width=0.5\textwidth]{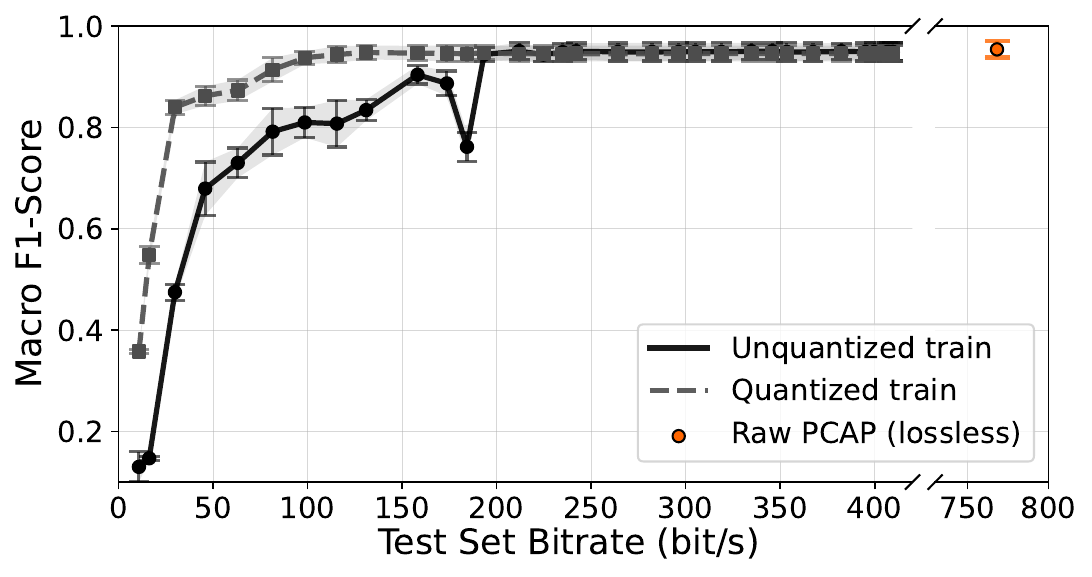}
        \label{fig:subfig1}}
        
        \subfigure[]{
        \includegraphics[width=0.5\textwidth]{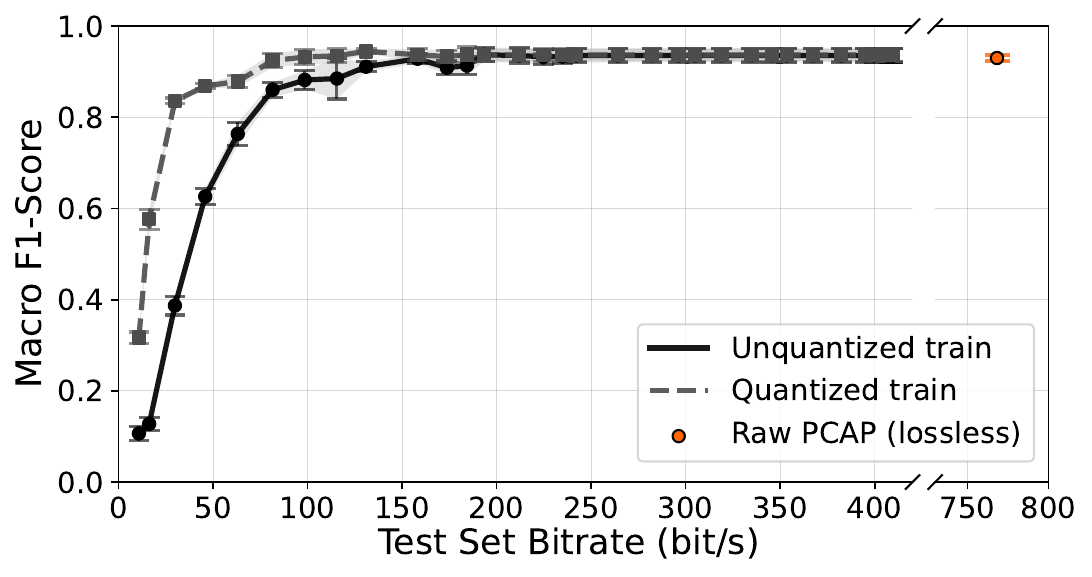}
        \label{fig:subfig2}}
    \caption{Impact of training set bit-rate on Macro F1 score for (a) device type and (b) individual device identification classification tasks. Results compare two approaches: models trained and tested on quantized data versus models trained on unquantized data but tested on quantized data.}
    \label{fig:quantization_f1_comparison}
\end{figure}

\subsection{Device Type Classification}
Figure~\ref{fig:subfig1} reports the impact of lossy feature compression on device type classification performance, showing the macro F1-score as a function of the average storage bitrate. Results are shown for classifiers trained on unquantized features and evaluated on quantized features, as well as for classifiers trained and tested on quantized representations. For reference, the performance achieved using lossless compression of raw \texttt{.pcap} traces is also indicated.

For both lossy configurations, classification performance improves with increasing bitrate, following a saturating trend. At very low bitrates, aggressive quantization leads to a marked degradation in performance, whereas moderate increases in bitrate rapidly restore classification accuracy.

When the classifier is trained and evaluated on quantized features, macro F1-scores above 0.9 are achieved at bitrates around 100~bit/s, and near-lossless performance is reached at approximately 150--200~bit/s. Beyond this point, further increases in bitrate yield only marginal improvements. In contrast, models trained on unquantized features converge more slowly and exhibit higher variability at intermediate bitrates.

\subsection{Individual Device Identification}
Overall, the performance trends observed for individual device identification are similar to those obtained for device type classification. When the classifier is trained and evaluated on quantized features, performance improves rapidly as the bitrate increases. Macro F1-scores above 0.85 are achieved at bitrates around 80--100~bit/s, while near-lossless performance (macro F1 $\approx$ 0.92--0.94) is reached at approximately 150--200~bit/s. Beyond this range, further increases in bitrate yield marginal performance gains, indicating that most device-specific discriminative information is preserved once a moderate feature precision is retained.

In contrast, models trained on unquantized features and evaluated on quantized representations exhibit slower convergence and higher variability at intermediate bitrates, consistently underperforming the quantized-training configuration. This behavior further confirms that exposing the classifier to quantization effects during training improves robustness to reduced feature precision.

\subsection{Discussion and Implications for IoT Forensics}
The results of this section show that storage strategies for Zigbee traffic impact not only classification accuracy, but also the set of forensic and analytical tasks that can be supported. Lossless storage of raw packet traces preserves complete information and enables the widest range of forensic analyses, including retrospective feature extraction and protocol-level inspection. However, even with state-of-the-art compression, this approach requires storage rates on the order of several hundred bit/s.
Feature-based storage provides a more compact representation at the cost of analytical flexibility. When lossy compression through feature quantization is adopted, storage requirements can be reduced by approximately 4--5$\times$ compared to lossless raw traffic storage, while maintaining near-lossless classification accuracy for both device type classification and individual device identification. Training classifiers directly on quantized features further improves robustness to reduced precision and enables the most storage-efficient operating points.
These gains come with functional limitations. Feature-based representations restrict subsequent analyses to the predefined feature set, preventing packet-level inspection and the extraction of alternative features. As a result, lossy feature compression is well suited for continuous monitoring and classification tasks at the edge, whereas lossless raw traffic storage remains preferable for scenarios requiring full forensic flexibility.

%% file: Chapters/conclusion_v2.tex
\section{Conclusions}
\label{sec:conclusion}
In this work, we introduced ZIOTP2025, a publicly available multi-topology dataset of Zigbee network traffic collected from 21 commercial smart home devices. With over 252,000 IEEE~802.15.4 frames captured under realistic interaction scenarios, the dataset enables systematic investigation of topology-dependent effects that are largely overlooked in existing IoT traffic datasets.

Using ZIOTP2025, we showed that state-of-the-art traffic classification methods achieve excellent performance under intra-topology evaluation, but suffer significant degradation when evaluated across different network topologies, particularly for fine-grained device identification. These results highlight network topology as a key factor limiting the generalizability of Zigbee traffic analysis models.

We also analyzed the trade-off between traffic storage requirements and classification accuracy. Our results show that lossy feature compression through quantization can reduce storage requirements by approximately 4--5$\times$ compared to lossless raw traffic storage, while preserving near-lossless classification accuracy. Overall, this work emphasizes the need for topology-aware analysis and storage-efficient designs in scalable IoT forensic systems.

Future work will focus on extending the dataset to additional topologies and device ecosystems, and on developing topology-aware analysis and adaptive storage strategies for scalable IoT forensic systems.